\documentclass[11pt,letterpaper]{article}
\usepackage[T1]{fontenc}
\usepackage{jheppub}
\usepackage{comment,xcolor}
\usepackage{amsmath}
\usepackage{hyperref}
\usepackage{mathtools}
\usepackage{amssymb}
\usepackage{amsfonts}
\usepackage{amsthm}
\usepackage{mathrsfs}
\usepackage[all]{xy}
\usepackage{tensor}
\usepackage{footnote}
\usepackage{graphicx} 
\usepackage[justification=raggedright,singlelinecheck=false]{caption}
\usepackage{subcaption}
\usepackage{cleveref}
\usepackage{booktabs}
\usepackage{enumitem}
\usepackage[export]{adjustbox}
\usepackage{soul}

\newcommand{\dd}{\mathrm{d}}
\newcommand{\beq}{\begin{equation}}
\newcommand{\eeq}{\end{equation}}
\numberwithin{equation}{section}

\newcommand{\ket}[1]{| #1 \rangle}
\newcommand{\bra}[1]{\langle #1 |}
\newcommand{\braket}[2]{\left\langle #1 \mid #2 \right\rangle}
\newcommand{\ketbra}[2]{\left|#1\right\rangle\!\!\left\langle #2\right|}

\def\eps{\varepsilon}

\def\N{\mathcal{N}}
\def\M{\mathcal{M}}
\def\O{\mathcal{O}}

\def\M{\mathcal{M}}

\def\H{\mathcal{H}}

\newcommand*{\tr}{\mathrm{Tr}}

\def\D{\mathcal{D}}

\def\E{\mathcal{E}}

\title{What exactly does Bekenstein bound?}
\author{Patrick Hayden, Jinzhao Wang}
\affiliation{\small \it Stanford Institute for Theoretical Physics, Stanford University, Stanford, CA 94305}
\emailAdd{phayden@stanford.edu}
\emailAdd{wang.jinzhao226@gmail.com}

\abstract{The Bekenstein bound posits a maximum entropy for matter with finite energy confined to a spatial region. It is often interpreted as a fundamental limit on the information that can be stored by physical objects. In this work, we test this interpretation by asking whether the Bekenstein bound imposes constraints on a channel's communication capacity, a context in which information can be given a mathematically rigorous and operationally meaningful definition. We study specifically the \emph{Unruh channel} that describes a stationary Alice exciting different species of free scalar fields to send information to an accelerating Bob, who is confined to a Rindler wedge and exposed to the noise of Unruh radiation. We show that the classical and quantum capacities of the Unruh channel obey the Bekenstein bound that pertains to the decoder Bob. In contrast, even at high temperatures, the Unruh channel can transmit a significant number of \emph{zero-bits}, which are quantum communication resources that can be used for quantum identification and many other primitive protocols. Therefore, unlike classical bits and qubits, zero-bits and their associated information processing capability are generally not constrained by the Bekenstein bound. However, we further show that when both the encoder and the decoder are restricted, the Bekenstein bound does constrain the channel capacities, including the zero-bit capacity.}

\begin{document}
\maketitle

\section{Introduction}\label{sec:intro}

Information-theoretic concepts have amply demonstrated their utility for understanding fundamental aspects of physics. One notable example is the Bekenstein bound~\cite{bekenstein1981universal}, originally proposed as a way to safeguard the second law of thermodynamics against violations by black holes~\cite{bekenstein1974generalized}. The bound asserts that the entropy $S$ of matter confined within a spatial region of size $R$ and energy $E$ is subject to a specific limit,
\begin{equation}\label{eq:bb}
    S\le \lambda RE\ .
\end{equation}
where $\lambda$ is some order-one constant independent of $G_N$. It is one of the first profound theoretical discoveries to identify an intrinsic connection between information and energy,\footnote{Another well-known result that bears a similar flavor is Landauer's principle~\cite{landauer1961irreversibility,bennett2003notes}, which posits that the minimal energy cost of erasing a bit is $k_B\ln 2$.} and inspired many subsequent developments. (See~\cite{bousso2020black} for an insightful review.) The Bekenstein bound serves as a compelling reminder that information, despite its abstract nature, is necessarily carried by physical systems.

The terms in the formula proposed by Bekenstein were ambiguously defined~\cite{page1982comment}. The proper formulation and validity of the bound remained elusive and critically debated (see \cite{bekenstein2005does} for a review), until Casini successfully proposed a precise version of equation \eqref{eq:bb} within the framework of quantum field theory~\cite{casini2008relative}. Casini made the observation that the relative entropy between the quantum state of matter, denoted $\rho$, and the vacuum state, denoted $\Omega$, can be expanded as the difference between an energy term and an entropy term,\footnote{Here some UV regularization is assumed to decompose the relative entropy into the two terms. }
\begin{equation}\label{eq:casini}
    S(\rho_B||\Omega_B)=\langle K_B\rangle_{\rho} - \delta S(\rho_B) \ge 0\ ,
\end{equation}
where the relative entropy is evaluated with respect to the observables supported on some region $B$; $\delta S(\rho_B)$ is the vacuum-subtracted von Neumann entropy of $\rho_B$; and $\langle K_B\rangle_{\rho}$ is the modular energy for which the vacuum energy is set to zero. (In general, the modular Hamiltonian $K_{\rho}$ for any positive operator $\rho$ is defined as $K_{\rho}:=-\log\rho$. We omit the subscript when the state is the vacuum.). Casini's entropy bound follows from the positivity of the relative entropy. 

The key insight of Casini's formulation is that the Bekenstein bound can be understood as an information-theoretic statement about state distinguishability, measured by the relative entropy in a hypothesis testing scenario. The bound arises from the fact that excited states are generally not perfectly distinguishable from the vacuum. The Bekenstein bound becomes interesting as it approaches saturation, when the state becomes indistinguishable from the vacuum locally within the region of interest.  The distinguishability can be improved by investing energy to better localize the distinguishing features of the states, as well as by providing more space to examine their differences.

However, even Casini's entropy bound does not perfectly capture Bekenstein's vision of information being bounded by a quantity that pertains to the spatial extent and energy of a region as in~\eqref{eq:bb}. In order to make Casini's entropy bound closer to~\eqref{eq:bb}, one needs to adopt the decomposition~\eqref{eq:casini}. There are issues with the decomposition both technically and conceptually. The technical issue is that the vacuum regularization is not obviously legitimate for arbitrary regions in a given QFT. Also, the vacuum-subtracted modular Hamiltonian does not admit the form $RE$ in general. Only in some special cases, c.f.~\cite{casini2008relative}, will the modular Hamiltonian have a form that resembles~\eqref{eq:bb}. The modular Hamiltonian of a Rindler wedge (specified by $x\ge 0$ on a $t=0$ slice), for instance, reads
\begin{equation}\label{eq:rindlerhamiltonian}
    K=2\pi\int\dd x^\perp\int_0^\infty\dd x\ x\ T_{00}(x)\ .
\end{equation}

One can certainly restrict to scenarios where the technical issues are resolvable. However, there is the further conceptual issue of \emph{operational meaning}. The von Neumann entropy characterizes the optimal rate at which the quantum state of matter can be compressed into a memory~\cite{schumacher1995quantum}. However, the operational interpretation becomes less apparent when the entropy is renormalized by subtracting the vacuum contribution. In particular, both sides of the inequality~\eqref{eq:casini} can be \emph{negative}, making the operational meaning even more obscure.\footnote{It is conceivable that the vacuum-subtracted entropy could be understood as a quantum conditional entropy, which admits negative values and characterizes operational tasks like state merging.} 

Casini's entropy bound is a widely accepted proven version of the Bekenstein bound, but it is not the only viable formulation~\cite{blanco2013localization,longo2018comment}. A general interpretation of the Bekenstein bound is that it represents a fundamental limit on the information content carried by matter~\cite{bekenstein2004black,bekenstein2005does}. Bekenstein himself also made efforts to generalize the scope of his original proposition to communication scenarios~\cite{bekenstein1981energy,bekenstein1984entropy,bekenstein1990quantum}.\footnote{See also~\cite{hod2007universal,carullo2021bekenstein,buoninfante2022bekenstein} for other applications.}  However, it is important to note that this interpretation remains folklore rather than established fact, mainly because different operational tasks lead to distinct realizations and characterizations of the notion of \emph{information}, among which the von Neumann entropy is just one instance.

We regard \emph{the Bekenstein bound} as a general principle that admits different formulations in different operational contexts. Our goal in this work is to test the scope of the Bekenstein bound as a bound on information from an operational perspective. We would like to know whether the existing proposals for the Bekenstein bound admit alternative formulations, such that they have clear operational meanings.

We follow Casini's idea that the Bekenstein bound, as we know it so far, is about state distinguishability. Besides the task of distinguishing two states as hypothesis testing, it is natural to consider distinguishing an ensemble of states. In Section~\ref{sec:review}, we first review a Bekenstein bound due to Bousso~\cite{bousso2017universal} on the \emph{accessible information} of an ensemble of states. It constrains the classical information that can be read out with measurements confined within a region. In addition to the operational meaning, the upshot of Bousso's bound is that the Holevo information is by itself UV finite, so no vacuum subtraction as in~\eqref{eq:casini} is needed. This bound applies to extracting classical information from a confined region, regardless of how the information is encoded in the first place.

\subsection{Our contributions}

Distinguishing quantum states is essential to classical communication over quantum channels. Does the Bekenstein bound apply universally to quantum communication? The natural framework to address this question, which incorporates both encoding and decoding, is quantum Shannon theory. Since the task of distinguishing states mostly pertains to the decoder/receiver, we first focus on bounding various channel capacities for channels that only restrain the decoding but not the encoding.  The relevant Bekenstein bound then depends on the energy of the codeword states perceived by the decoder and his spatial constraint.

The capacities describe the amount of classical and quantum information transmittable per channel use in the limit of many uses of identical and independent distributed (i.i.d.) channels. Again, a virtue of channel capacities, as compared to the von Neumann entropy, is that they are naturally UV-finite and operationally meaningful. 

In Section~\ref{sec:unruh}, instead of answering the question in full generality, we study the concrete model of free scalar fields in Rindler spacetime. We consider communication from a stationary Alice, who sees the entire space, and an accelerating Bob, who only sees a portion of space. Alice encodes her messages by exciting a particle from one of several distinct species, and Bob receives these signals and tries to decode Alice's message. The fact that Bob's decoding operations are constrained within the Rindler wedge is equivalent to the presence of Unruh radiation~\cite{fulling1973nonuniqueness,davies1975scalar,unruh1976notes} that adds noise to this communication channel. We shall refer to the channel as the \emph{Unruh channel}. 

We evaluate its various channel capacities for transmitting classical and quantum information. Our findings reveal that both the quantum and the classical capacities of the communication channel is bounded by the Bekenstein quantity $\langle K_B\rangle=\beta E$, where $\beta$ is the inverse temperature that Bob measures and $E$ is the energy of the message signals he measures.  

However, we find that the entanglement-assisted (classical) capacity is \textit{not} constrained by the Bekenstein bound. Notably, even in scenarios where Bob is infinitely accelerating, so the unassisted classical capacity is negligibly small and $\beta E \rightarrow  0$, entanglement assistance allows the transmission of an amount of classical information growing without bound as the number of species grows. This violation, however, is well expected because the physical medium that carries the auxiliary Bell pairs is not accounted for in the Bekenstein bound. Nonetheless, the capacity boost due to entanglement assistance hints that the channel still robustly transmits some exotic form of quantum information even at high temperatures.\\

In Section~\ref{sec:zerobit}, we discuss the main result of our work, a \emph{genuine violation} of the Bekenstein bound. This considerable capacity boost due to entanglement assistance can be attributed to specific quantum communication resources known as \emph{zero-bits}, first studied in the context of quantum identification by Hayden-Winter~\cite{hayden2012weak} and later formalized by Hayden-Penington~\cite{hayden2020approximate}. Zero-bits serve as minimal substitutes for classical bits (cbits) or qubits in various primitive quantum information processing protocols such as teleportation and dense coding. They are well-defined and useful, however, even in the absence of entanglement assistance. Although zero-bits are less powerful than cbits or qubits, they are more robust against noise. 

We show that the Unruh channel possesses a large zero-bit capacity regardless of the level of background noise caused by the Unruh radiation. Moreover, this capacity increases indefinitely as the number of particle species accessible to Alice grows. In this regard, unlike cbits and qubits, there is no Bekenstein bound that universally constrains zero-bits and their information-processing capabilities. In Appendix~\ref{sec:alpha}, we further show that $\alpha$-bits, which interpolate between the zero-bits and the qubits, also violate the Bekenstein bound.\\

What if one also restrains the encoder's spatial domain and energy budget? Does it help constrain the zero-bit capacity? In previous studies of the Bekenstein bound, such subtleties did not arise because people sought the Bekenstein bound that works for any state. In other words, it was implicitly assumed that the encoding is not restricted at all. In the communication setting, however, it is physically reasonable to consider physical restrictions on the encoder. Then it is sensible to expect that the constraints on the encoder and the decoder should both be manifest in the Bekenstein bound. 

Using another version of the Bekenstein bound due to Blanco-Casini~\cite{blanco2013localization}, we show in Section~\ref{sec:encoder} that any quantum channel with appropriately restricted encoding and decoding does obey the Bekenstein bound. Importantly, the zero-bit capacity is also bounded. Combined with an earlier result due to Page reviewed in Appendix~\ref{sec:counterexample}, we thus show that it is necessary for both the encoder and the decoder to be restricted in order for the Bekenstein bound to constrain all the channel capacities considered here. This is our second main result.

In Section~\ref{sec:conclusion}, we close with some remarks.

\section{Motivation: The Bekenstein bound on Accessible Information}\label{sec:review}
To see how to generalize the state distinguishability task to communication, let us consider the task of Bob extracting classical information from an ensemble of quantum states encoded by Alice. Let the random variable describing Alice's message be $X$, which is distributed according to the law $p_X$. For each message $x\in \mathcal{X}$, Alice assigns a codeword $\rho^x$ which is a density operator in the operator algebra of the QFT. Bob has access to any measurement operations supported in a region $B$, and tries to measure the codewords and figure out the classical information stored by Alice. Note that Alice's encoding was not spatially restricted. Let the random variable describing Bob's outcome be $Y$. The task is to maximize the mutual information between $X$ and $Y$ over decoding measurements. We define the \emph{accessible information} as the maximal mutual information,
\begin{equation}
    I_\mathrm{acc}(B,\rho):= \max_{\{\Pi_y\}} I(X:Y)_{P_{XY}},\quad P_{XY}:=\sum_{x\in\mathcal{X}}\sum_{y\in\mathcal{X}}\tr\ \Pi_y\rho_B^x \ketbra{x}{x}_X\otimes\ketbra{y}{y}_Y \ ,
\end{equation}
where $\{\Pi_y\}$ is a POVM set, and we use $\rho$ as a shorthand for the ensemble $\{p_x,\rho_B^x\}$.

We would like to know whether $I_\mathrm{acc}(B,\rho)$ obeys the Bekenstein bound. This quantity does not have a general closed-form formula, but there is a useful upper bound known as the Holevo information~\cite{holevo1973bounds},
\begin{equation}\label{eq:holevoinfo}
    I_\mathrm{acc}(B,\rho) \le \chi(\{p_x,\rho_B^x\}):=\sum_{x\in\mathcal{X}} p_x S(\rho_B^x||\bar\rho_B)\ ,
\end{equation}
where $\bar\rho_B:=\sum_{x\in\mathcal{X}} p_x\rho_B^x$. The relative entropy here is well-defined in QFT using Tomita-Takesaki theory and Araki's definition. The bound will be saturated if and only if the ensemble consists of commuting codewords, such as an orthogonal set.

Bousso derived a bound on the Holevo information in~\cite{bousso2017universal}. We now review his results. Note the following identity\footnote{This identity holds generally for states on von Neumann algebras.} for the Holevo information for any ensemble of states~\cite{donald1987further},
\begin{equation}
\chi(\{p_x,\rho^x_B\})=\sum_{x\in\mathcal{X}} p_x S(\rho^x_B||\Omega_B)-S(\bar\rho_B||\Omega_B)\ ,
\end{equation}
where the reference state $\Omega$ is chosen to be the vacuum.\footnote{One can pick any appropriate state as the reference depending on the context. For example, for a black hole in equilibrium, the Hartle-Hawking state could be chosen instead.}

On the RHS, we have the relative entropy between the mixture $\bar\rho$ and the vacuum $\Omega$ over the region $B$, and its positivity is what Casini identified as the Bekenstein bound~\eqref{eq:casini}. It follows that
\begin{equation}\label{eq:bound_on_chi}
I_\mathrm{acc}(B,\rho)\le\chi(\{p_x,\rho^x_B\})\le \sum_{x\in\mathcal{X}} p_x S(\rho^x_B||\Omega_B)\ .
\end{equation}

The Holevo information and thus the accessible information are bounded by the average relative entropy between the codewords and the vacuum in the region $B$. This is a Casini bound on the accessible information. Expanding both sides we find
\begin{equation} \label{eq:bound_on_chi_regularized}
S(\rho)-\sum_{x\in\mathcal{X}}p_x S(\rho^x_B)\le \langle K_B\rangle_{\bar\rho}-\sum_{x\in\mathcal{X}} p_x S(\rho^x_B)=\sum_{x\in\mathcal{X}} p_x\left(\langle K_B\rangle_{\rho^x_B}- S(\rho^x_B)\right)\ .
\end{equation}
Therefore, it is the same as Casini’s bound \eqref{eq:casini} up to the regularization. While the vacuum-subtracted energy may be natural, vacuum-subtracted entropy is more ad hoc. The regularized quantities in \eqref{eq:bound_on_chi_regularized}, in contrast, are always well-defined as relative entropies. 
While Casini's entropy bound $\langle K_B\rangle_\rho$ is linear in the quantum state $\rho$, \eqref{eq:bound_on_chi_regularized} is linear in $p_X$, which describes different sources of classical information. As compared to Casini's bound on entropy, the Holevo information is operationally meaningful and nonnegative. Bousso further considered signals that are classical enough that the codewords are more entropic than the vacuum,\footnote{In any relativistic quantum field theory, the vacuum is infinitely entangled, so it has a UV-divergent entanglement entropy. Here the inequality must be understood as after subtracting the same UV-divergent piece from both sides.} 
\begin{equation}\label{eq:more_entropic}
    S(\rho_B^x)\ge S(\Omega_B)\ .
\end{equation}
This further bounds~\eqref{eq:bound_on_chi} with $\langle K_B\rangle_\rho$ as in Casini's entropy bound.

We see from Bousso's result that it is desirable to consider information measures that pertain to communication instead of the von Neumann entropy. The upshot is that this measure is operationally meaningful and intrinsically UV finite.


\section{The Bekenstein bound on the Unruh channel capacities}\label{sec:unruh}

The accessible information pertains to classical information. To understand the scope of the Bekenstein bound for quantum information or other operational notions of information decodable from a region, we need to adopt the general framework of quantum Shannon theory. The theory was developed for the primary purpose of analyzing the capacities of quantum communication channels. We can incorporate various energy and spatial constraints in the encoding and decoding into the description of a quantum channel, and use the Shannon theory toolkit to look for appropriate Bekenstein bounds for their channel capacities. 

In this language, the Bekenstein bound we obtained in the previous section can in turn be applied to bound the unfortunately named ``one-shot''\footnote{The terminology is misleading. The one-shot capacity is more properly called the product state capacity of the channel. It is the maximum rate at which bits can be communicated over many uses of the channel provided codewords are not entangled between successive uses of the channel.} classical capacity of a quantum channel. This is because the Holevo information of the output codewords optimized over the input upper bounds the one-shot classical capacity of a quantum channel with outputs restricted to the region $B$.\footnote{More precisely, with a single use of the channel, the rate achievable by a channel $\N$ with error $\eps$ is given by the hypothesis testing relative entropy~\cite{wang2012one}. The converse is upper bounded by $[\chi(\N)+h(\eps)]/(1-\eps)$. Up to the binary entropy term $h(\eps)$ which is controlled by the precision required, the one-shot classical capacity is bounded by the Holevo information.}

Judging from earlier efforts~\cite{bekenstein1981energy,bekenstein1984entropy,bekenstein1990quantum,bousso2017universal}, it is challenging to deduce precise bounds on communication in generality. Instead, we study a particular channel, known as the Unruh channel, in Rindler space. This channel models a communication scenario where the sender Alice has access to the entire Minkowski space whereas the receiver Bob only has access to the right Rindler wedge. We will grant Alice the ability to encode her message in an unconstrained way, but Bob's decoding is constrained within the Rindler wedge. A priori, one would expect that the standard Bekenstein bound in Rindler space puts a limit on the capacities of the Unruh channel. As we will see, however, that is not always true. The Bekenstein bound can be violated for some more exotic forms of quantum communication.

The same setup was used by Marolf-Minic-Ross (MMR)~\cite{marolf2004notes,marolf2005few} to resolve the \emph{species problem}~\cite{page1982comment}. The concerning issue was the possibility of increasing the entropy without a corresponding increase in energy by introducing additional particle species, which suggested a possible violation of the Bekenstein bound. The work of MMR was an important precursor to Casini's entropy bound. It will prove to be valuable to revisit their model and explore a different set of information-theoretic measures to test the validity of the Bekenstein bound.

Incidentally, the Unruh channel was analyzed by Bradler-Hayden-Panangaden (BHP)~\cite{bradler2012quantum} in studying communication in Rindler space. (See also~\cite{bradler2009private,bradler2010conjugate}). Their work revealed several useful technical properties of the Unruh channel that will prove beneficial for our calculations. While BHP evaluated the quantum capacity, their primary focus was not the Bekenstein bound. In our study, we will build upon their results and investigate whether the Bekenstein bound is respected by various capacities of the Unruh channel.

\subsection{Communication with the Unruh channel}
\begin{figure}
    \centering
    \includegraphics[width=.45\linewidth]{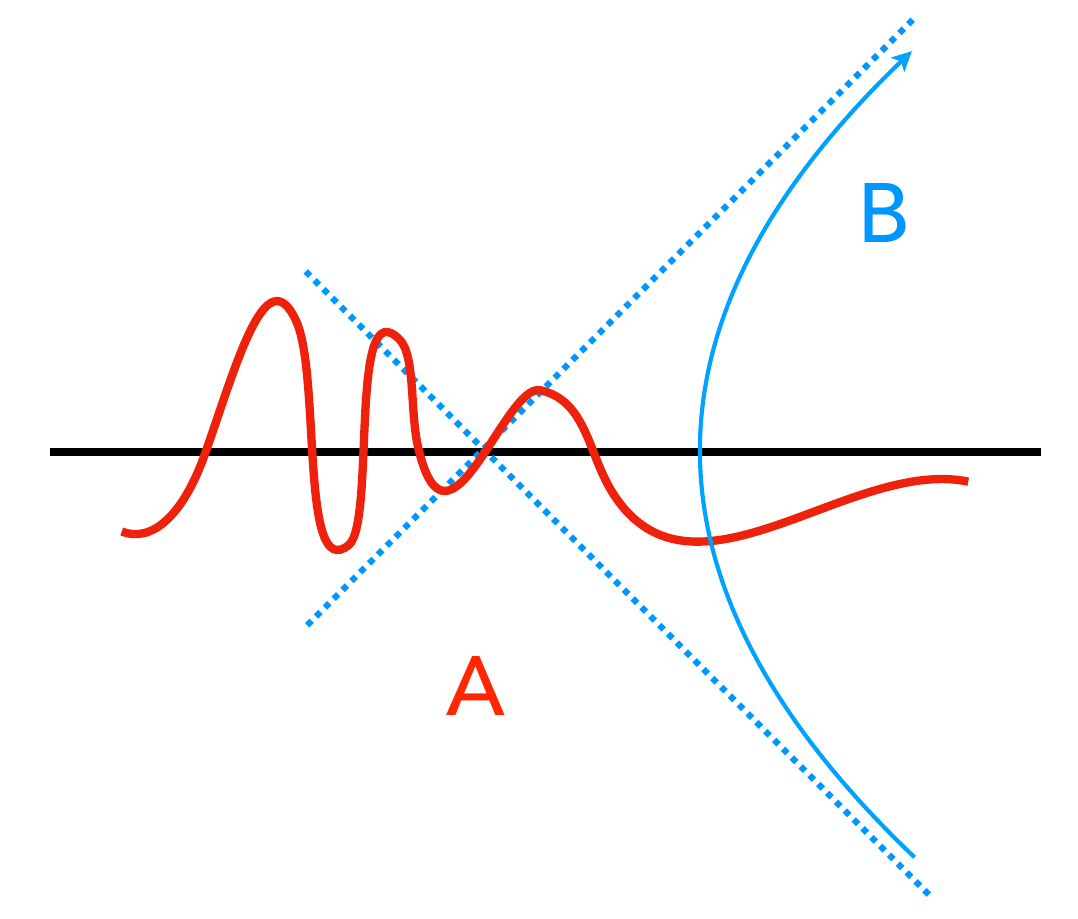}
    \caption{\textbf{Unruh channel.} The global agent Alice prepares the codewords in an Unruh mode and the accelerating Bob receives them with noise. }
    \label{fig:unruh_channel}
\end{figure}

We consider free scalar fields of $d$ distinct species in Minkowski spacetime.  Alice is a stationary observer who sees the entire spacetime. Bob is a right Rindler mover with acceleration $a$. He experiences Unruh radiation at the temperature of $\beta=2\pi/a$, which also measures the proper distance to the Rindler horizon. Alice communicates with Bob by exciting modes in QFT, and Bob receives the noisy signals disrupted by the Unruh radiation. (cf. Fig.~\ref{fig:unruh_channel} for an illustration.)

For simplicity, we restrict Alice's encoding to single-particle states in the \emph{right Unruh mode} $U^R_{i,\beta,\omega}$~\cite{unruh1976notes}, which are analytic continuations of the right Rindler modes $R^R_{i,\beta,\omega}$ that are labeled by the temperature $\beta$ and the frequency $\omega$ measured in Bob's frame. The index $i\in [d]$ labels the particle species. Using a Bogoliubov transformation, we can write the creation operator of $U^R_{i,\beta,\omega}$, denoted $a^\dagger_{iR}$, in terms of the left and right Rindler creation and annihilation operators of $R^L_{i,\beta,\omega}$ and $R^R_{i,\beta,\omega}$, denoted $b_{iL}^\dagger$ and $b_{iR}^\dagger$,
\begin{equation}\label{eq:bogoliubov}
    a^\dagger_{iR} = (1-e^{-\beta \omega})^{-\frac12}(b_{iR}^\dagger - e^{-\beta \omega/2} b_{iL})\ .
\end{equation}

Let $\H_A$ denote the $d$-dimensional subspace of these single-particle states in distinct species sectors. We restrict Alice's encoding to the single excitation subspace via
\begin{equation}
    V^{(d)}:  \H_A\to \H,\quad  \ket{i}\mapsto\ket{0}_1\otimes\cdots\otimes a^\dagger_{iR}\ket{0}_i\otimes\cdots\otimes\ket{0}_d \ ,
\end{equation}
where $\H$ denotes the Hilbert space of quantum fields, $a^\dagger_{iR}$ denotes the creation operator for $U^R_{i,\beta,\omega}$ and the superscript~$(d)$ denotes the input dimension. Note that the Unruh modes share the same vacuum as the Minkowski modes, so the vacuum state in the above equation denotes the Minkowski/Unruh vacuum.

The modes propagate to Bob, who perceives the particle content differently than Alice. The Minkowski/Unruh vacuum can be represented using the Rindler Fock basis appropriate to Bob's perspective,
\begin{equation}
    \ket{0}=\bigotimes_{i=1}^d \ket{0}_i = \bigotimes_{i=1}^d\bigotimes_{R_v} \sum_{N_i^v=0}^\infty\sqrt{1-e^{-\beta\omega_v}} e^{-\frac{\beta}{2}\omega_v N_i^v}\ket{N^v_i,N^v_i}\ ,
\end{equation}
where $i$ labels the particle species and $v$ labels the Rindler modes, and $\ket{N^v_i,N^v_i}$ denotes the Fock states of the mode $R_v$ in the $i^\mathrm{th}$-particle sector. 

The other modes are simply not relevant in this calculation, so we might as well ignore them. Let each $\ket{\Omega}_i$ be the \emph{projected vacuum} that only involves a single pair of modes $(R^L_{i,\beta,\omega}, R^R_{i,\beta,\omega})$,
\begin{equation}\label{eq:projectedvac}
    \ket{\Omega}_i = P\ket{0}_i = \sum_{N_i=0}^\infty\sqrt{1-e^{-\beta\omega}} e^{-\frac{\beta}{2}\omega N_i}\ket{N_i,N_i},\quad \ket{\Omega}=\bigotimes_{i=1}^d\ket{\Omega}_i \ .
\end{equation}
where $(N_i,N_i)$ is the number of particles in this pair of Rindler modes $(R^L_{i,\beta,\omega}, R^R_{i,\beta,\omega})$.

The remaining vacuum entanglement is among the Fock states of this mode, without the infinite tensor product among different modes. Therefore, we can write down the density matrix of the reduced state in the right (or left) Rindler wedge, which does not exist for the full vacuum. 
\begin{equation}\label{eq:vacuum}
    \Omega=\bigotimes_{i=1}^d(1-e^{-\beta \omega})e^{-N_i\beta \omega} =(1-e^{-\beta \omega})^de^{-N\beta \omega}\ , \quad N:=\sum_{i=1}^d N_i=\sum_{i=1}^d b^\dagger_{iR}b_{iR}\ .
\end{equation} 

We label Bob's right Rindler wedge as $B$. Since Bob is immersed in the Unruh radiation, the message he receives from Alice is degraded by the noise.  This thermal noise fundamentally originates from tracing out the left Rindler wedge $\bar B$. We refer to this noisy communication channel as the \emph{Unruh channel},
\begin{equation}
    \N_{d,\beta,\omega}:= \tr_{\bar B}(V^{(d)}\,\cdot\,V^{(d)}): \mathcal{P}(\H_A)\to \mathcal{P}(\H_B)\ ,
\end{equation}
We have identified the Hilbert space of the right Rindler wedge as Bob's Hilbert space $\H_B$.\footnote{\label{ft:factorization} Note that the quantum field theory Hilbert space $\H$ is generally nonfactorizable. However, when restricted to the subspace of particles in a single mode, the restricted Hilbert space $\H$ does factorize into $\H=\H_L\otimes\H_R$.}

We will study the Unruh channels, parameterized by their input dimension $d$, or equivalently, the number of particle species that Alice has access to. They are further parameterized by the inverse temperature that determines the noise and the energy  $\omega$ of the Unruh mode. We shall henceforth omit the superscripts and subscripts in $\N_{d,\beta,\omega}$ for notational clarity. \\

Before proceeding, let us briefly comment on the physicality of this communication channel. Note that we did not model the time evolution of the mode or give Bob any time window for decoding. Therefore we are describing a simplified and idealized communication scenario, where there Bob has infinite computational power to decode the message in a short amount of time. This is acceptable so long as we are testing the validity of the Bekenstein bound from the purely information-theoretic perspective.

Unruh modes are often used to approximate Minkowski modes in the so-called single-mode approximation~\cite{alsing2003teleportation}, which is sometimes problematic~\cite{bruschi2010unruh}. We assume Alice can prepare the Unruh modes directly rather than using them as an approximation. Also, we restrict Alice to encode the message with single-particle excitations. These assumptions are convenient because their Bogoliubov transformation to the Rindler modes is easier to analyze than the one for Minkowski modes. This technical assumption helps to facilitate channel capacity calculations. However, we should mention other the Unruh modes are less regular than the Minkowski modes. For an inertial observer, they are non-monochromatic, ill-localized, and rapidly oscillatoring near the Rindler horizon. Hence, they are much more difficult to prepare physically. 

Another caveat is that we would need Alice to operate nonlocally because we are considering single particle states that cannot be prepared with quantum channels that act locally (without post-selection).  Likewise, for Bob to decode the message in the Rindler wedge, he might also need to implement measurements/operations that are nonlocal across the wedge. In this sense, the Unruh channel describes an idealized communication scenario that might not be physically realizable. 

However, to study the questions of principle we are concerned with here, the Unruh channel is still a useful model for testing certain operational aspects of the Bekenstein bound. We grant Alice nonlocal operations to prepare globally distinguishable codewords while constraining Bob to the Rindler wedge. We would like to know if the Bekenstein bound limits how much information Bob can decode.\footnote{There is a related study of information transfer using Unruh modes~\cite{martin2012fundamental}, where general mixtures of the left and right Unruh modes are studied. The analysis is not carried out in the framework of Shannon theory, but similar quantities like Holevo information and coherent information are computed. It would be interesting to extend our analysis to those cases as well.}

\subsection{A lightning review of channel capacities}
We now review some basics about channel capacities. They precisely characterize the maximal number of cbits, qubits that can be reliably transmitted through the channel. In particular, we will study the classical capacity, quantum capacity, and entanglement-assisted capacities.  See~\cite{wilde2013quantum} for a comprehensive review.

In the case of classical communication over quantum channels, the message is modeled as a discrete random variable $X$ valued in the alphabet $\mathcal{X}$. Alice encodes her message in a joint quantum state $\rho_x$ that is input to a channel $\M:\mathcal{P}(\H_A)\to \mathcal{P}(\H_B)$, where we use $\mathcal{P}(\H)$ to denote positive operators on the Hilbert space~$\H$. Then Bob decodes Alice's transmitted message by measuring the output state using his POVM $\{E_x\}_{x\in\mathcal{X}}$ to generate the output random variable $Y$. We say the protocol $(\{\rho_x\},\{E_x\})$ is a $(k,\eps)$-code if $|\mathcal{X}|=k$ and the transmission error is bounded by $\eps$,
\begin{equation}
 \mathrm{Pr}(X\neq Y|X=x) = \tr(1-E_x)\M(\rho_x)\leq \eps, \forall x\ .
\end{equation}
Let~$k_\eps$ denote the largest message size given an error $\eps$,
\begin{equation}
    k_\eps(\M):=\max\{k\in\mathbb{N}\ |\ \exists\,\,\text{a}\,\, (k,\eps)\text{-code}\,\,\mathrm{for}\,\,\M\}\ .
\end{equation}
We say $r$ is an achievable rate for $\M$ if $\forall\eps>0$,  $\log k_\eps(\M^{\otimes n})\ge rn$  for large enough $n$. The classical capacity $C(\M)$ is defined as the supremum over achievable rates,
\begin{equation}
    C(\M):=\lim_{\eps\to 0}\lim_{n\to \infty}\frac1n\lfloor\log k_\eps(\M^{\otimes n})\rfloor\ .
\end{equation}
The famous Holevo-Schumacher-Westmoreland theorem~\cite{holevo1998capacity,schumacher1996sending,schumacher1997sending} demonstrates that the classical capacity is given by the regularized Holevo capacity,
\begin{equation}
    C(\M)=\lim_{n\to\infty}\frac1n\chi(\M^{\otimes n})\ ,
\end{equation}
where the Holevo capacity is defined as 
\begin{equation}\label{eq:holevo}
    \chi(\M):=\max_{\{p_x,\psi_x\}} \left[S\left(\sum_x p_x\M(\psi_x)\right) - \sum_x p_xS(\M(\psi_x))\right]\ ,
\end{equation}
where the maximization is over the encoded ensemble of pure states $\{\psi_x\}_{x\in\mathcal{X}}$ and random variables $X$.\footnote{As in the case for the Holevo information~\eqref{eq:holevoinfo}, the Holevo capacity has an alternative formulation in terms of the relative entropy~\cite{schumacher2001optimal,schumacher2002relative}, $\chi(\M) = \min_{\sigma}\max_{\rho} S(\rho\circ\M^\dagger||\sigma)$. This Heisenberg-picture formula is manifestly well defined for subalgebras in QFT.} The regularized Holevo capacity is a \emph{multi-letter} formula that is generally difficult to evaluate.

Now consider Alice sending quantum information to Bob instead of classical messages. We demand that the entanglement between the quantum message and any reference purification be preserved by the communication. This extra requirement implies that the quantum capacity can never exceed the classical capacity. Let Alice's encoding channel be $\E:\mathcal{P}(\H_X)\to \mathcal{P}(\H_A)$, and Bob's decoding channel be $\D:\mathcal{P}(\H_B)\to \mathcal{P}(\H_Y)$ we say the protocol $(\E,\D)$ is a $(k,\eps)$-quantum code if $\dim \H_X = k$ and
\begin{equation}
    ||\D\circ\M\circ\E-\mathcal{I}||_\diamond\le \eps\ ,
\end{equation}
where the diamond norm of a channel $\M_{A\to B}$ is defined as $||\M||_\diamond:=\max_{\psi_{AR}}||\mathcal{I}\otimes\M(\psi)||_1$. As for the classical capacity, we can then define the quantum capacity as the supremum over all achievable rates,
\begin{equation}
    Q(\M):=\lim_{\eps\to 0}\lim_{n\to \infty}\frac1n\lfloor\log k^q_\eps(\M^{\otimes n})\rfloor\ ,
\end{equation}
where $k^q_\eps:=\max\{k\in\mathbb{N}\ |\ \exists\,\,\text{a}\,\, (k,\eps)\text{-quantum}\,\,\mathrm{code}\,\,\mathrm{for}\,\,\M\}$.

The coherent information $I_c(\M)$ for a channel $\M$ is defined as
\begin{equation}\label{eq:coherentinfo}
    I_c(\M):=\max_{\psi\in\H_{A}\otimes\H_{A'}}I(A\rangle B)_{\M(\psi)}:=\max_{\psi\in\H_{A}\otimes\H_{A'}} S(B)_{\M(\psi)}-S(A'B)_{\M(\psi)}\ ,
\end{equation}
where $\psi\in\H_A\otimes\H_{A'}$, $\M(\psi)\in\H_{A'}\otimes\H_{B}$, and $A'$ is an auxiliary system isomorphic to $A$ that acts as a reference system to help capture the quantum information in $A$.

The Lloyd-Shor-Devetak theorem~\cite{lloyd1997capacity,shor2002quantum,devetak2005private} shows that the quantum capacity is given by the regularized coherent information,
\begin{equation}
    Q(\M):=\lim_{n\to\infty}\frac1n I_c(\M^{\otimes n})\ .
\end{equation}

We are also interested in the classical and quantum capacities of a channel when Alice and Bob are given \emph{free entanglement} as a resource. In that setting, Alice encodes her messages into her half of a set of shared Bell pairs and sends the encoded halves to Bob through the channel. Bob will try to decode Alice's messages with the help of his half of the Bell pairs. Since entanglement can only help them achieve better communication rates, the entanglement-assisted capacities are generally higher than without the assistance. The capacities are defined in a similar way, so we shall not repeat the details. We will use two useful facts here.  Bennett et al.~\cite{bennett1999entanglement,bennett2002entanglement} showed that the entanglement-assisted classical capacity is always given by a \emph{single-letter} formula that doesn't involve multiple copies,
\begin{equation}\label{eq:C_E}
    C_E(\M):=\max_{\psi\in\H_{A}\otimes\H_{A'}} I(A':B)_{\M(\psi)}:=\max_{\psi\in\H_{A}\otimes\H_{A'}} \left(S(A')_\psi+S(B)_{\mathcal{I}\otimes\M(\psi)}-S(A'B)_{\mathcal{I}\otimes\M(\psi)}\right)\ .
\end{equation}
We can also rewrite the formula as
\begin{equation}\label{eq:eacapacity}
    C_E(\M):=\max_{\psi\in\H_{A}\otimes\H_{A'}} \left(S(A)_\psi+I(A\rangle B)_{\M(\psi)}\right)\ .
\end{equation}
The second fact is that with entanglement assistance, the classical capacity is twice the quantum capacity, $C_E=2Q_E$, so it is sufficient to focus here on the classical capacity.\\

BHP has studied a variant of the Unruh channel, where messages are encoded in distinguishable modes instead of in different sectors of particle species~\cite{bradler2012quantum}. Despite the physical difference, these two versions of Unruh channels are mathematically identical. (See Definition~11 in BHP for a more detailed description of the Unruh channel.) Using group-theoretic tools, BHP established several useful properties of the Unruh channel. We list a few that will be useful later. 

\begin{itemize}
    \item \emph{Covariance.} A covariant channel has its input and output transform covariantly under the same unitary group. More precisely, we say that a channel $\M:\mathcal{P}(\H_A)\to \mathcal{P}(\H_B)$ is $G$-covariant with respect to representations $r_1:G\to \mathrm{GL}(\H_A), r_2:G\to \mathrm{GL}(\H_B)$, if $\M(r_1(g)\rho\, r_1(g)^\dagger)=r_2(g)\M(\rho) r_2(g)^\dagger,\ \forall g\in G,\rho\in \mathcal{P}(\H_A)$. The claim is that the Unruh channel is $\mathrm{SU}(d)$-covariant with respect to the fundamental representation $U(g)$ in the input space and the isometrically mapped representation $VU(g)V^\dagger$ in the output space. 
    
    To see why, consider an arbitrary unitary $U$ in the fundamental representation and a rotated input state $U\rho U^\dagger$. We have $\mathcal{V}(U\rho U^\dagger)= (VUV^\dagger)\mathcal{V}(\rho)(VUV^\dagger)^\dagger.$ The encoded unitary $VUV^\dagger$ only acts by shuffling the tensor factors of different species, without acting within each factor at all. Hence, $VUV^\dagger$ commutes with tracing out the left Rindler wedge, which acts on each individual factor. The $\mathrm{SU}(d)$-covariance follows.  
    \item \emph{Degradability.} The Unruh channel is transposable, that is, up to a transpose ($\mathcal{T}$), the complement channel is equal to the primary channel concatenated with some other channel,
    \begin{equation}
       \exists \E, \ \mathrm{s.t.}\,\,\, \mathcal{T}\circ\N^c = \E\circ\N\ .
    \end{equation}
    Physically, (transpose) degradability means the information that leaks to the environment can always be simulated from the primary channel. In this sense, it is never the case that ``more information'' flows to the left wedge. Since quantum information cannot be cloned, the complementary channel of a degradable channel has zero quantum capacity. This is quite intuitive for the Unruh channel as the Unruh mode is analytically extended from the corresponding Rindler mode. The degrading map is literally preparing the analytical extension from the output Rindler mode.  
    
    The upshot is that although the quantum capacity is generally superadditive, it is additive for degradable channels~\cite{devetak2005capacity}. We therefore have a single-letter formula,
\begin{equation}\label{eq:qcapacity}
    Q(\N)=\max_{\psi\in\H_{A}\otimes\H_{A'}}I(A\rangle B)_{\N(\psi)}\ .
\end{equation}
    \item \emph{Hadamard.} The Unruh channel is a Hadamard channel~\cite{bradler2011infinite},\footnote{The channel action in the Choi representation takes the form of a Hadamard product.} meaning that its complement channel is entanglement-breaking.\footnote{Entanglement-breaking channels output separable states for any inputs entangled with a reference~\cite{horodecki2003entanglement}. They have additive classical capacities and zero quantum capacity. Here $\N^c$ being entanglement-breaking follows from the fact that its transpose $\mathcal{T}\circ\N^c$ is also a quantum channel.} A Hadamard channel is a special instance of a degradable channel:\footnote{Hence, in fact, the Unruh channel is both degradable and transpose degradable.} Its degrading map can be decomposed as two maps, where the first one maps the channel output to a classical description and the second one prepares the complementary output according to the classical description. 
    
    The upshot is that although the classical capacity is generally superadditive~\cite{hastings2009superadditivity}, it is additive for Hadamard channels~\cite{king2005properties}. We thus have a single-letter formula,
\begin{equation}\label{eq:ccapacity}
    C(\N)=\chi(\N)\ .
\end{equation}
\end{itemize}

\subsection{Bounding the output entropy}\label{sec:bekenstein}

As mentioned earlier, free scalar fields of multiple species in Rindler space were used by MMR to resolve the species problem. In terms of the setup we have introduced, they computed the output entropy of the Unruh channel for maximally mixed states of increasing dimension corresponding to the species number $d$. They found that the entropy saturated as $d$ was increased, just as the Bekenstein bound itself approached saturation, thereby avoiding the expected violation of the bound. The entropy saturation effect provided a resolution to the species problem.

Historically, MMR's calculation was an important precursor to Casini's general proof of the Bekenstein bound in Rindler space. In order to find the appropriate Bekenstein bound for the Unruh channel, it is instructive to review the main result of MMR in light of Casini's entropy bound. We measure all the entropies and capacities in units of nats instead of bits and use the natural logarithm by default.

Consider the encoded maximally mixed state $\pi:=\frac{1}{d}\sum_{i=1}^d\ketbra{i}{i}$,
\begin{equation}
    \N(\pi)=\frac1d\sum_{i=1}^d\rho_i =\frac{e^{\beta \omega}-1}{d}N\Omega\ ,
\end{equation}
where $N$ is the total number operator, and $\Omega$ is the right Rindler wedge reduced density matrix of the projected vacuum~\eqref{eq:vacuum}.

Its entropy was evaluated approximately by MMR~\cite{marolf2004notes}, then numerically by Marolf~\cite{marolf2005few}, and later analytically by Casini and BHP~\cite{casini2008relative,bradler2012quantum}. Here we quote the formula from BHP:
\begin{multline}\label{eq:entropy_pi}
    H(d):=S(\N(\pi))=S(\Omega)+\log d-\log (e^{\beta\omega}-1)+\frac{  \beta\omega}{1-e^{-\beta\omega}}\\-\frac{(1-e^{-\beta\omega})^{d+1}}{d}\sum_{k=1}^\infty\binom{d+k-1}{k}e^{-\beta\omega(k-1)}k\log k e^{-\beta\omega(k-1)}\ .
\end{multline}
We will refer to this entropy quantity several times, so it will be convenient to define it as a function of the input dimension, $H(d)$. In the formula, $S(\Omega)$ denotes the von Neumann entropy of the vacuum (right) density operator~\eqref{eq:vacuum},
\begin{equation}
    S(\Omega) = d\left(\frac{\beta \omega  }{e^{\beta \omega}-1}-\log(1-e^{-\beta \omega})\right)\ .
\end{equation}
We can make the following approximations for the entropy of $\N(\pi)$,
\begin{equation}
    \delta S := S(\N(\pi))-S(\Omega) \stackrel{\beta \omega\gg 1}{\approx} \begin{cases}
        \log d ,  \quad\log d\le \beta \omega\\
        \beta \omega  , \quad\quad \log d\ge \beta \omega\ .
    \end{cases}
\end{equation}
The vacuum-subtracted entropy $\delta S$ captures the entropy of the logical information.  We see that $\delta S$ increases with $\log d$ until it saturates at $\beta \omega  $. $S$ is thus constrained by the Bekenstein bound. That is the main takeaway from MMR's work. 

To work out the Bekenstein bound \`a la Casini, we evaluate the relative entropy~\cite{casini2008relative}, 
\begin{equation}\label{eq:casini_unruh}
    S(\N(\pi)||\Omega)= \frac{\beta\omega}{1-e^{-\beta \omega}} - \delta S\ ,
\end{equation}
and it follows that
\begin{equation}
    \delta S\le \frac{\beta\omega}{1-e^{-\beta \omega}}\ ,
\end{equation}
which is now an exact bound that doesn't involve the approximation $\beta \omega\gg 1$ as above. The RHS is the modular energy~\eqref{eq:rindlerhamiltonian},
\begin{equation}
    \langle K_B\rangle_{\N(\pi)}=\beta\omega \langle N \rangle_{\N(\pi)}=\frac{\beta\omega}{1-e^{-\beta \omega}}=:\beta E.
\end{equation}
We will use the shorthand notation $\langle K_B\rangle$ to denote this bound. This is unambiguous because the value is the same for any input state $\rho$, $\langle K_B\rangle_{\N(\pi)}=\langle K_B\rangle_{\N(\rho)}$.

Thus, we obtain 
\begin{equation}
    \delta S\le \langle K_B\rangle=\beta E\ .
\end{equation}
This can also be understood as the statement that the (coarse-grained) thermal entropy is larger than the (fine-grained) information-theoretic entropy. 

Moreover, this Bekenstein bound applies to all states. Because the von Neumann entropy function is concave and the Unruh channel is $\mathrm{SU}(d)$-covariant, the maximal output entropy is achieved for the maximally mixed input state. We therefore find that 
\begin{equation}
    \forall\rho,\quad S(\N(\rho))\le S(\N(\pi))\le  \langle K_B\rangle= \beta E\ .
\end{equation}
 
We have established that $\langle K_B\rangle$ is the appropriate Bekenstein bound for the Unruh channel as far as the output entropy is concerned. Note that this bound only pertains to restrictions on the decoder Bob, because we do not restrict the encoder Alice at all.  We shall now study the extent to which the Bekenstein bound actually constrains the ability of Alice to communicate with Bob.

\subsection{Bounding the channel capacities}

Now let us evaluate these capacities for the Unruh channel and see if they too are constrained by the Bekenstein bound. Unlike the von Neumann entropy, the channel capacities are often expressed as differences between two entropic quantities so the UV divergences cancel out. We therefore do not need to impose the vacuum subtraction by hand. We shall compare the results against both Casini's entropy bound of the vacuum-subtracted modular energy $\langle K_B\rangle=\beta E$ as well as the general bound we had in Section~\ref{sec:review}.\\

\emph{Classical capacity.} To find the maximizer for the Holevo capacity~\eqref{eq:holevo}, note that the second term is independent of the message random variable $X$ and the ensemble $\{\psi_x\}$, i.e. $\sum_x p_x S(\N(\psi_x))=S(\N(\psi_x))$, because $\N$ is covariant so $S(\N(\psi_x))=S(\N(\psi_y)),\ \forall x,y$.

We only need to maximize the first term $S\left(\sum_x p_x\N(\psi_x)\right)$. As we have argued, because of the channel covariance and the concavity of von Neumann entropy, the maximizer is the maximally mixed state $\pi$. Likewise, the Holevo capacity is maximized by a uniformly distributed ensemble over computational basis states, $\{1/d,\ketbra{i}{i}\}_{i=1}^d$. We get
\begin{equation}\label{eq:chi}
    \chi(\N) = S(\N(\pi)) - \frac1d\sum_i S(\N(\ketbra{i}{i})) =S(\N(\pi))-S(\N(\ketbra{i}{i}))\ ,
\end{equation}
where the second equality follows from covariance, $S(\N(\ketbra{i}{i}))=S(\N(\ketbra{j}{j})),\ \forall i,j$.

The first term is $H(d)$, given by~\eqref{eq:entropy_pi}.
For the second term, we have
\begin{equation}\label{eq:channel_output}
    \N(\ketbra{i}{i})=(e^{\beta \omega}-1)N_i\Omega\ ,
\end{equation}
where $N_i$ is the number operator for the $i^\mathrm{th}$-particle. The entropy evaluates to
\begin{equation}
    S(\N(\ketbra{i}{i})) = H(1)+\frac{d-1}{d}S(\Omega)\ ,
\end{equation}
and so we find
\begin{equation}
    \chi(\N) = H(d)-H(1)-\frac{d-1}{d}S(\Omega)\ .
\end{equation}

To estimate its value, let us rewrite it as
\begin{equation}
    \chi(\N) = H(d)-S(\Omega)+S(\Omega)/d-H(1)= \delta S+\left[S(\Omega)/d-H(1)\right]\ .
\end{equation}
The second term is some order-one quantity $-\gamma   \le\left[S(\Omega)/d-H(1)\right]\le 0$ where  $\gamma\approx 0.577$ is Euler's constant. (See Appendix~\ref{sec:estimates}.) So we have
\begin{equation}
    \chi(\N) \le \delta S\le \beta E   .
\end{equation}
We conclude that the classical capacity obeys the Bekenstein bound. Interestingly, even in the infinite temperature limit $(\beta\to 0)$, the classical capacity is small but nonzero,
\begin{equation}
    \chi(\N)\to (1-\gamma)\approx 0.61   .
\end{equation}
Note that this does not violate the Bekenstein bound as $\beta E\to 1$ in this limit. We can also verify Bousso's bound on the Holevo information~\eqref{eq:bound_on_chi}. It gives a slightly tighter bound than the Bekenstein bound $\beta E$ because of~\eqref{eq:more_entropic} and
\begin{equation}
    S(\N(\ketbra{i}{i}))-S(\Omega) = H(1) - S(\Omega)/d\ge 0\ .
\end{equation}

\begin{figure}
    \centering
    \includegraphics[width=0.95\linewidth]{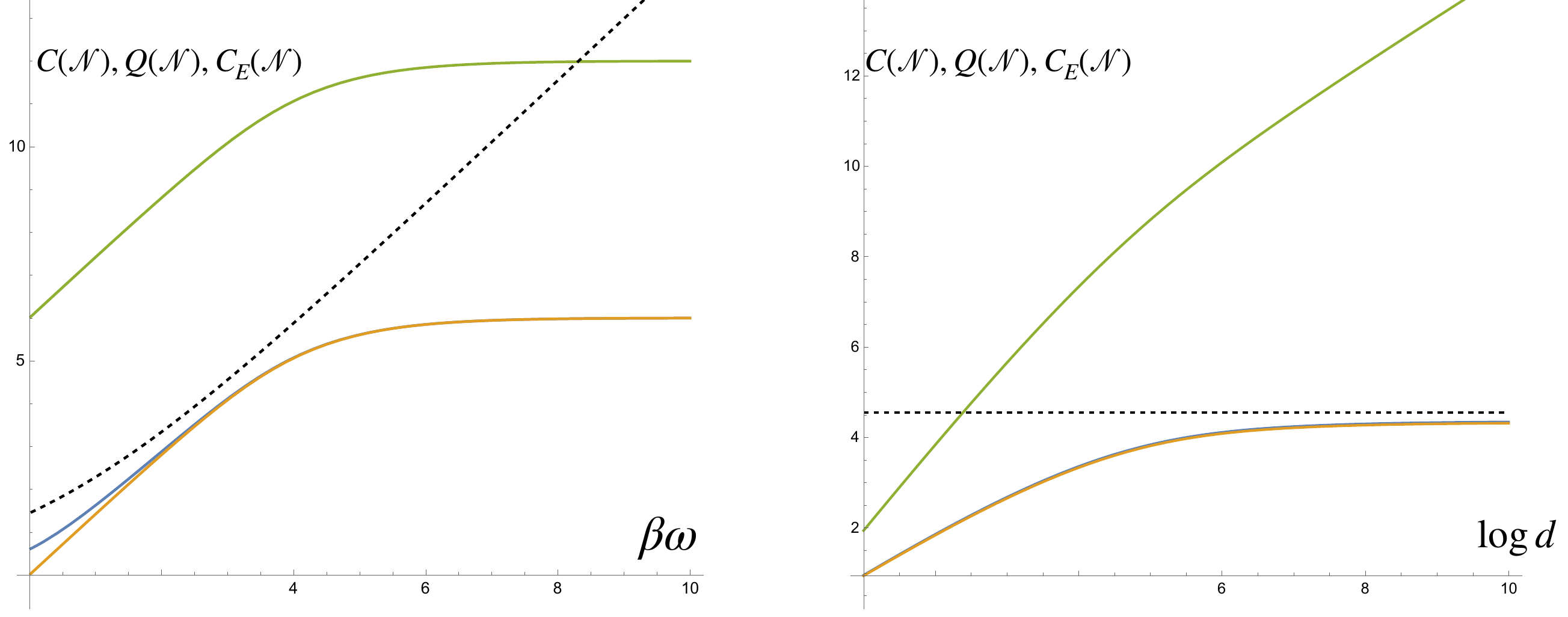}
    \caption{\textbf{Capacities of the Unruh channel.} The classical capacity $C(\N)$ in blue and the quantum capacity $Q(\N)$ in orange plotted gainst the channel parameters $\beta\omega$ and $\log d$ respectively. They both satisfy the Bekenstein bound $\langle K_B\rangle =\beta\omega/(1-e^{-\beta\omega})$ plotted as the dashed line.  However, the entanglement-assisted classical capacity $C_E(\N)$ in green does violate the Beksntein bound. On the left figure, $C(\N)$ and $Q(\N)$ asymptote to the maximum $\log d$ and $C_E(\N)$ asymptotes to the maximum $2\log d$. We have $C(\N)\approx Q(\N)\approx\min\{\log d,\beta E\}$. }
    \label{fig:cq_capacity}
\end{figure}

\emph{Quantum capacity.} To evaluate the quantum capacity, we need to calculate the maximal coherent information~\eqref{eq:qcapacity}. Since the Unruh channel is both covariant and transpose-degradable, the maximum is once again achieved by the maximally mixed state $\pi$. The quantum capacity is simply given by $S(\N(\pi))-S(\N^c(\pi))$. This has been evaluated by BHP. The first term is given in~\eqref{eq:entropy_pi} and the second term is given in Appendix~\ref{sec:estimates}. The quantum capacity is~\cite{bradler2012quantum}
\begin{equation}\label{eq:qc}
    Q(\N)=S(\N(\pi))-S(\N^c(\pi))=\frac{1}{d(1-e^{-\beta\omega})^{d+1}}\sum_{k=1}^\infty \binom{d+k-1}{k}k e^{-\beta\omega(k-1)}\log\frac{d+k-1}{k}\ .
\end{equation}
We can rewrite the expression as
\begin{equation}
    Q(\N)=\delta S + \left[S(\Omega)-S(\N^c(\pi))\right]\ ,
\end{equation}
where the second term in the bracket is negative and lower bounded by $-1$ (see Appendix~\ref{sec:estimates}), so we have
\begin{equation}
    Q(\N)\le\delta S\le \beta E  = \frac{\beta\omega}{1-e^{-\beta\omega}}  \ .
\end{equation}
We conclude that the quantum capacity obeys the Bekenstein bound. In fact, we already know this because the quantum capacity is upper bounded by the classical capacity, which obeys the Bekenstein bound. Unlike the classical capacity, $Q(\N)$ drops to zero as $\beta\to 0$. Note that $C(\N)$ is not much larger than $Q(\N)$ as they are both less than $\delta S$ and within one bit. We have 
\begin{equation}
    Q(\N)\le C(\N)\le Q(\N)+1\ .
\end{equation}

\emph{Entanglement-assisted capacity.} For the entanglement-assisted classical capacity~\eqref{eq:eacapacity}, it is clear that both terms are maximized by the maximally entangled state. We thus obtain
\begin{equation}
    C_E(\N)=\log d+Q(\N) \ge\log d+\delta S-1\ .
\end{equation}

This is a rather simple calculation, but physically a quite striking result. Recall that the classical capacity of the Unruh channel is almost negligible in the infinite temperature limit. Nonetheless, with entanglement assistance, the capacity is at least $\log d$, which is extensive in the number of input qubits. Indeed, in that limit, the entanglement-assisted capacity of the channel almost matches that of a noiseless classical channel on $d$ letters despite it being almost useless for communication without entanglement assistance!

Entanglement assistance can \emph{at most} boost the capacity of any channel up by $\log d$, i.e.,  $C_E\le\log d+C$. Since $C(\N)-Q(\N)\le 1$, the entanglement assistance is \emph{nearly maximal}. Such a big boost in classical capacity was not known in the literature for any physically motivated channels.\footnote{Bennett el al showed that the ratio $C_E/C$ can be large for the depolarizing channel in the large noise limit~\cite{bennett1999entanglement}, but both capacities also tend to zero. The case of the Unruh channel is more surprising because both capacities remain finite.} The same kind of capacity boost also exists for the entanglement-assisted quantum capacity, $Q_E=C_E/2\ge \frac12\log d$. However, for quantum capacities, such a separation is more common. For example, the noiseless classical bit channel has zero quantum capacity, but with entanglement assistance, we have $Q_E=\frac12\log d$.  \\

Note that $C_E(\N)$ is not bounded above by $\langle K_B\rangle=\beta E$. Bob ends up receiving more bits than the Bekenstein bound allows. This happens because we have allowed Bob to decode the message using his share of the additional entanglement resources that the Bekenstein bound does not take into account. If we forbid Bob from using his halves of the Bell pairs in decoding, then Alice running her dense coding protocol with her shares is of no value for enhancing the classical communication. It follows that the best achievable rate will just be the classical capacity $C(\N)$ and the Bekenstein bound will be respected.  

Therefore, if we allow Bob to decode with his share of the auxiliary system, we also need to account for the contributions from the Bell pairs, despite the fact that Alice never acted on the auxiliary system. To ensure that the entanglement is a free resource between Alice and Bob, we can assume that the Bell pairs are kept in some cavities held by Alice and Bob.\footnote{The cavities can be made to preserve the prepared entanglement under acceleration~\cite{schutzhold2005comment,downes2011entangling,alsing2012observer}.} The degrees of freedom are described by some auxiliary Hilbert space $\H'_A\otimes\H'_B$ independent from the free fields in Rindler space. Since the protocol consumes at most $\log d$ Bell states per channel use, we then need to increase the Bekenstein bound by $\log d$.\footnote{A quick way to see this $\log d$ increase is to look at Casini's entropy bound. When~$\log d$ Bell pairs $\Phi^{\otimes \log d}_{AB}$ are included, the relevant bound is $S(\rho\otimes\Phi^{\otimes \log d}_B||\Omega\otimes\Phi^{\otimes \log d}_B)\ge 0$, which is equivalent to $\langle K_B\rangle_\rho+\log d\ge\delta S(\rho)+\log d$. }

\section{No Bekenstein bound for zero-bits}\label{sec:zerobit}

A puzzle concerning the large capacity boost from entanglement assistance remains. We need to bear in mind that free entanglement alone does not allow Alice to communicate with Bob. On the other hand, the Unruh channel becomes so noisy at high temperatures that no qubits and only a negligible fraction of a classical bit can be sent, regardless of how many particle species are at Alice's disposal. What is then the communication resource transmitted by the Unruh channel that allows Alice and Bob to utilize the entanglement and achieve the capacity of at least $\log d$ ?

It is instructive to see explicitly how well we can distinguish two noisy codewords $\N(\ketbra{i}{i})$, $\N(\ketbra{j}{j})$ for any orthogonal $\ket{i}, \ket{j}$. We measure the distinguishability using fidelity. The smaller the fidelity is, the more distinguishable the two states are. Using \eqref{eq:channel_output}, we obtain
\begin{equation}\label{eq:fidelity_codewords1}
    F(\N(\ketbra{i}{i}),\N(\ketbra{j}{j}))= (1-e^{-\beta \omega})^2(e^{\beta \omega}-1)\operatorname{Li}_{-\frac12}(e^{-\beta \omega})^2\stackrel{\beta \omega\gg 1}{\approx} e^{-\beta \omega}\ ,
\end{equation}
where we approximate the polylogarithm function for a small argument by the leading term in the expansion. We conclude that the fidelity is suppressed exponentially in the Bekenstein quantity.\footnote{Note that this error, however small, is not zero, so the error can accumulate to set a limit on how well one can distinguish a large ensemble of states. It is therefore expected that it is increasingly difficult to distinguish all the members of a larger ensemble of noisy codewords. This intuitively explains why the classical capacity also obeys the Bekenstein bound. } We therefore say the Unruh channel is able to preserve geometry when $\beta \omega\gg 1$.

Hayden-Winter realized that geometry preservation implies that the noisy channel still maintains some amount of quantum coherence that can be utilized~\cite{hayden2012weak}. This feature can be operationalized in the task of \emph{quantum identification} (QID), where Bob is supposed to simulate measurements that test whether the state sent by Alice is an arbitrary pure state. More precisely, we say that $(\E,D)$ is an $(k,\eps)$-QID code for a channel $\M:\mathcal{P}(\H_A)\to \mathcal{P}(\H_B)$, if  for any projective measurement $\{\ketbra{\varphi}{\varphi},1-\ketbra{\varphi}{\varphi}\}$ ($\varphi\in\H_X$) that Bob wants to simulate on his output end, there exists a binary POVM $\{D_\varphi,1-D_\varphi\}$ that Bob can implement such that for any $k$-dimensional input state $\psi\in\H_X$ that Alice encodes using $\E:\mathcal{P}(\H_X)\to \mathcal{P}(\H_A)$, Bob can approximately simulate the ideal measurement effect on $\psi$,
\begin{equation}
    \bigg\vert\tr (\M\circ\E(\psi)\cdot D_\varphi) - |\braket{\varphi}{\psi}|^2\bigg\vert\le \eps\ .
\end{equation}
It is the universality that makes the task challenging: The same QID code must work for all Alice message states and all Bob projective measurements.
The claim is that a channel that preserves geometry well can also be used for quantum identification, and vice versa.

Hayden-Penington (HP)~\cite{hayden2020approximate} later formalized geometry preservation by referring to this quantum information transmitted by the channel as a \emph{zero-dit} with error $\eps$, where $d$ is the input dimension. A zero-dit is a one-shot notion associated with an error quantifier. It is more elegant and conceptually useful to make a further abstraction by introducing the \emph{zero-bit} as an asymptotic communication resource. We can use QID to operationally define zero-bits. We say a channel transmits $\log_2 k$ zero-bits\footnote{Note that while we measure capacities in nats, we switch to $\log_2$ when we refer to the number of zero-bits. Otherwise, it would be $\log d$ ``zero-nats''.} if for any $\eps>0$, there exists a $(k^n,\eps)$-QID code for $\N^{\otimes n}$ at sufficiently large $n$.

Let us use the Unruh channel as a concrete example. To obtain a larger input dimension, we use the channel $n$ times, but still only want to decode a two-dimensional subspace. The repetition helps to exponentially suppress the error. By encoding into a random subspace of the typical subspace of the input, it is possible to use $\N^{\otimes n}$ to send a zero-d$^{n-o(n)}$it with error decreasing exponentially with $n$~\cite{hayden2012weak,hayden2020approximate}.\footnote{The optimal zero-bit code is especially simple in this case because the coherent information is positive at all temperatures, eliminating the need for some complications found in the general case~\cite{hayden2012weak,hayden2020approximate}.} In the asymptotic limit $n\to\infty$, the error vanishes and we say the Unruh channel sends $\log_2 d$ \emph{zero-bits} per channel use. As compared to the zero-dit, the notion of zero-bit gets rid of any error quantifiers and therefore proves to be very useful as a building block for communication protocols. 

\begin{figure}
    \centering
    \includegraphics[width=.95\linewidth]{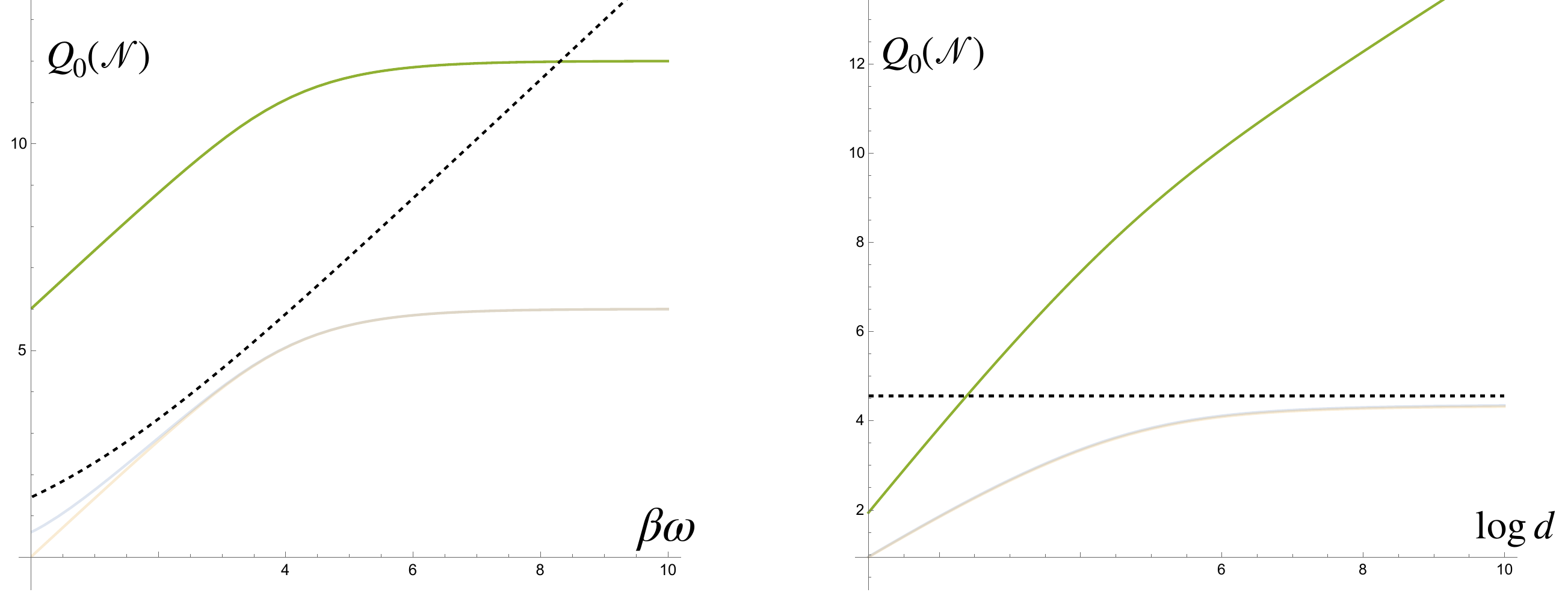}
    \caption{\textbf{Zero-bit capacity of the Unruh channel.} We plot $Q_0(\N)$ in green against $\beta\omega$ and $\log d$ respectively. They do not obey the Bekenstein bound. The opaque capacity curves of $C(\N), Q(\N)$ are there for comparison. The right figure shows that $Q_0(\N)$ grows unboundedly with $\log_2 d$, so zero-bits have the species problem. }
    \label{fig:0capacity}
\end{figure}

The argument above implies that the Unruh channel can at least achieve a communication rate of $\log_2 d$  zero-bits per channel use.  We can define the \emph{zero-bit capacity} for a channel as the supremum of achievable rates, denoted $Q_0$. This capacity is also known as the \emph{quantum identification capacity}, which measures the optimal asymptotic achievable rate at which the Unruh channel can support QID with vanishing error. HP showed that the zero-bit capacity admits a general formula that single-letterizes for any (transpose) degradable channel, such as the Unruh channel. The single-letter formula reads
\begin{equation}\label{eq:zerocapacity}
    Q_0(\N) =  \sup_{\psi\in\H_A\otimes\H_{A'}}\ [I(A':B)_{\N(\psi)} \quad \mathrm{s.t.} \quad I(A\rangle B)_{\N(\psi)}>0]\ ,
\end{equation}
where $I(A':B)_{\N(\psi)}$ is the channel mutual information, as in the formula for $C_E$~\eqref{eq:eacapacity}, and the coherent information is as in the formula for $Q$~\eqref{eq:qcapacity}. Note that $C_E\ge Q_0$ because they have the same formula except that $Q_0$ has an extra condition demanding strictly positive coherent information. For the Unruh channel, both $C_E(\N)$ and $Q(\N)$ are evaluated at a maximally entangled state $\psi=\Phi_{AA'}$, so we know that supremum for $Q_0$ is achieved also at the same state provided the strict positivity constraint is satisfied for this state. Fortunately, it is indeed satisfied because we have shown that $Q(\N)>0$ for all parameters $\beta,\omega$ of the channel. We therefore have (see Fig~\ref{fig:0capacity})
\begin{equation}\label{eq:zerocapacity_unruh}
    Q_0(\N) = C_E(\N) = \log_2 d+Q(\N)\ .
\end{equation}

Let us elaborate on this result. The zero-bit may sound like a mildly useful communication resource. However, besides its use in quantum identification, we can substitute the classical/quantum bits for zero-bits in many primitive quantum information processing protocols, such as teleportation, state merging, dense coding, entanglement distillation, and so on~\cite{hayden2020approximate}. They are also the minimal resources needed for these protocols, and so become readily useful in scenarios when the standard protocols are too resource-intensive. The relevant protocol in the context of classical communication is the zero-bit version of dense coding. Standard dense coding achieves the following resource inequality,
\begin{equation}\label{eq:dense_coding}
    1\; \mathrm{qubit} + 2\; \mathrm{ebits}\ge 2\; \mathrm{cbits}\ ,
\end{equation}
which means that one qubit and two Bell pairs (ebits) are sufficient to simulate the communication resource of two classical bits (cbits). However, it turns out that qubits are not necessary for this task, and we can replace them with two zero-bits. There is even a variant of the dense coding protocol that can operate in a noisy communication channel that is not capable of sending qubits provided it can send zero-bits, to help animate the Bell pairs by turning them into classical communication~\cite{hayden2020approximate}. We obtain
\begin{equation}\label{eq:zero_dense_coding}
   1\; \text{zero-bit} + 1\; \mathrm{ebit} \ge 1\; \mathrm{cbit}\ .
\end{equation}
This can be viewed as a tightening of the dense coding resource inequality~\eqref{eq:dense_coding}.\footnote{To make these identities rather than inequalities, one needs to substitute a \emph{coherent bit} on the RHS~\cite{harrow2004coherent}, which is stronger than a cbit.} We review the protocol in Appendix~\ref{sec:densecoding}.

Zero-bits are hence directly responsible for the capacity boost of entanglement-assisted classical communication. In the high temperature limit, the quantum capacity tends to zero, so we have \emph{no qubits} available for the standard textbook dense coding protocol~\eqref{eq:dense_coding}. However, we can still use the zero-bit version~\eqref{eq:zero_dense_coding} because the Unruh channel can send $\log_2 d$ zero-bits. With free entanglement, the entanglement-assisted classical capacity $C_E$ is at least $\log d$. Recall that generally $C_E\ge Q_0$, but this is saturated for the Unruh channel~\eqref{eq:zerocapacity_unruh}. This means no more bits can be sent once all the zero-bits are consumed. 
We thus identify the essential ingredient for the capacity boost in a channel that is seemingly too noisy to be useful. 

Since $Q_0$ increases unboundedly with the number of species $d$, we conclude that \emph{zero-bits are not constrained by the Bekenstein bound}. Bob can then further process and transmit zero-bits in various quantum information processing tasks.\footnote{Note that zero-bits can be ``turned’’ into cbits and qubits when combined with other resources such as ebits, such as dense coding or teleportation, and then the Bekenstein bound is obeyed by these so-obtained cbits and qubits.} 

One might think that there could be some alternative version of the Bekenstein bound that is not exactly $\langle K_B\rangle$ such that the zero-bit capacity can be constrained. This is not possible because zero-bits have the species problem as shown in the right figure in~\ref{fig:0capacity}. For any finite bound determined by the physical parameters that describe the mode, it can always be violated by the zero-bit capacity at a sufficiently large species number $d$. In contrast, the von Neumann entropy, classical and quantum capacities do not have the species problem. Therefore, there is simply no Bekenstein bound for zero-bits.

We re-emphasize that our claim is not in tension with Casini's bound in Rindler space, as we are not making a statement regarding the (regularized) von Neumann entropy but rather the zero-bit channel capacity. Our result instead challenges the folklore interpretation of the Bekenstein bound as a universal bound on the information, which in this case are the zero-bits transmitted from Alice to Bob. 

One caveat that might undermine our result is that this violation is an artifact of choosing the Unruh mode of various species as codewords. We make this choice for the convenience of additivity. Since the Unruh mode is singular at the horizon, it is technically possible that the Unruh mode could be fine-tuned for this zero-bit species problem. To check the robustness of our example, one needs to adopt and perhaps generalize the continuity statements concerning approximate additive channels~\cite{sutter2017approximate,leditzky2018approaches}, which claim that channels close in diamond norm to an additive one have capacities close to the additive counterparts. 

Such robustness is physically expected. It would be very surprising if this large violation could be canceled or even just alleviated when one regularizes the mode at the Rindler horizon, as we do not see any reasons of why this singular behavior at the horizon could help enhance Bob's decoding. We shall however leave the robustness analysis to future works.

\section{Restraining the encoder}\label{sec:encoder}
Is there a way to impose a Bekenstein bound on the zero-bits? We have shown that this is impossible if we only restrain the decoder. What if we also restrain the encoder and factor the restrictions in the Bekenstein bound? That is, we will consider channels that have certain spatially and energetically constrained inputs.  It turns out that one can indeed constrain the number of zero-bits transmitted when both the encoder and the decoder are restrained appropriately. To show this, we use another version of the Bekenstein bound proved by Blanco-Casini~\cite{blanco2013localization}, and consider a scenario in which we can interpret it operationally as a bound on channel capacities.

\begin{figure}
     \centering
     \begin{subfigure}[b]{0.48\textwidth}
         \centering
         \includegraphics[width=\textwidth]{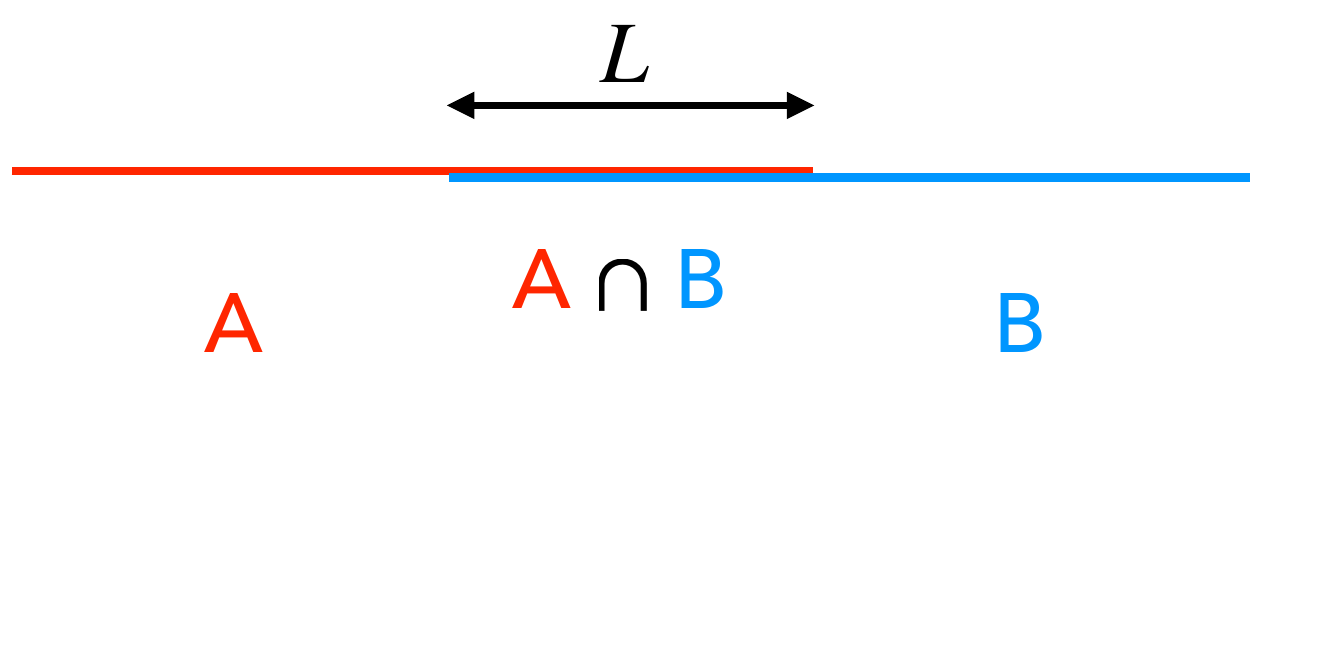}
         \caption{\centering Blanco-Casini bound}
         \label{fig:casini2}
     \end{subfigure}
     \hfill
     \begin{subfigure}[b]{0.48\textwidth}
         \centering
         \includegraphics[width=\textwidth]{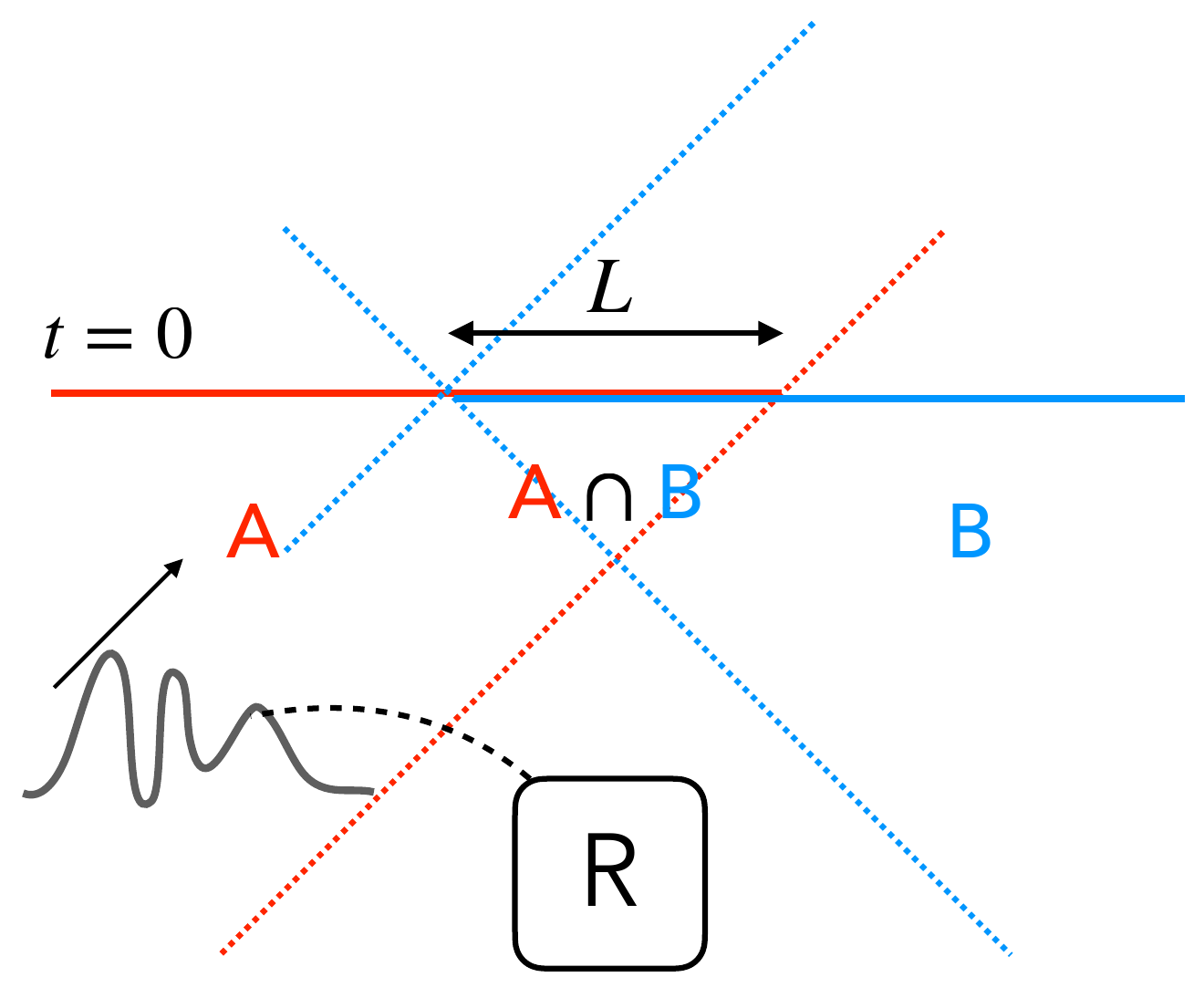}
         \caption{\centering Constrained Alice and Bob}
         \label{fig:constrained_Alice}
     \end{subfigure}
        \caption{Communication with constraints on both the encoder and the decoder. Fig.~\ref{fig:casini2} depicts two overlapping half-spaces for which the Blanco-Casini bound applies. Fig.~\ref{fig:constrained_Alice} depicts the spatial constraints for the encoder Alice and the decoder Bob.}
        \label{fig:result2}
\end{figure}

Consider a spatial slice shown in Fig.~\ref{fig:casini2}. $A$ and $B$ denote two half-spaces that overlap over a strip region $A\cap B$. Consider now the two relevant full modular Hamiltonians $\hat K_A:=K_A-K_{B\setminus A}$ and $\hat K_B:=K_B-K_{A\setminus B}$, where the modular Hamiltonians are defined for the reduced vacuum states on the corresponding spatial regions $A$ and $B$. It follows from strong subadditivty that $\hat K_A+\hat K_B$ is a positive operator. For its expectation value, Blanco-Casini strengthened it with an entropic lower bound. For any mixed state $\rho$,\footnote{This bound is trivial, i.e., equal to zero, for pure states.}
\begin{equation}
    \langle\hat K_A+\hat K_B\rangle_\rho\ge S(B)_\rho-S(B\setminus A)_\rho+S(A)_\rho-S(A\setminus B)_\rho\ .
\end{equation}
This bound follows directly from the monotonicity of relative entropy applied twice for shrinking $B$ to $B\setminus A$ and expanding $A\setminus B$ to $A$.

One can be more specific about the LHS. Let the width of $A\cap B$ be $L$. It follows that $\hat K_A+\hat K_B=2\pi LH$, where $H=\int\dd x^\perp\int_{-\infty}^\infty\dd x \ T_{00}$ is the total Hamiltonian. For any mixed state $\rho_{B\bar B}$ that is purified by $\rho_{B\bar B R}$, we can rewrite the entropy difference as
\begin{equation}
    I(A\cap B:R|B\setminus A)=I(B:R)_\rho-I(B\setminus A:R)_\rho\le 2\pi L\langle H\rangle_\rho\ .
\end{equation}
As compared to the more well-known bound of Casini~\eqref{eq:casini}, this version of the Bekenstein bound has the upshot that both sides are non-negative, and the bound has manifestly the canonical form~\eqref{eq:bb}. However, the operational meaning of this conditional mutual information has not been elucidated by Blanco-Casini.\footnote{One possible interpretation is that the (classical or quantum) communication resource that needed for the task of redistributing the system $A\cap B$ from $A$ to $B$~\cite{devetak2008exact}. }

Let us consider a communication scenario in which Alice encodes her messages by acting locally in $A$. This is illustrated in Fig.~\ref{fig:constrained_Alice}. Considering the time evolution, we let Alice encode her state in some half space in the past and demand that the future light cone intersects Bob's $t=0$ slice at the subregion $A\cap B$. Therefore, Alice's encoding region and Bob's decoding region overlap over a strip region $A\cap B$ at $t=0$. Compared to the Unruh channel, we have now constrained Alice's encoding operations. The receiver Bob still sees the right Rindler wedge defined by the half-space $B$. 

Consider now any quantum channel $\M_{A\to B}$ that satisfies the above conditions. Let $R$ be an auxiliary reference that entangles with Alice's message supported on $A$.  The channel takes any finite-dimensional input state $\psi\in\H_A\otimes\H_R$ and encodes the quantum message in $A$ into some field-theoretic modes supported within the past lightcone of Alice. The state evolves and then gets partially traced over $A\setminus B$ at $t=0$.\footnote{Our analysis only focuses on the $t=0$ slice w.l.o.g.. This is because any other half-Cauchy slices in Bob's wedge doesn't contain the $A\setminus B$ part of the message, which is all that is lost by Bob at the $t=0$ slice.} It follows from locality that for any input state $\psi$, $I(B\setminus A:R)_{\M(\psi)}=0$. Then the Blanco-Casini bound gives
\begin{equation}
    I(B:R)_{\M(\psi)}\le 2\pi L\langle H\rangle_{\M(\psi)},\quad \forall\psi\in\H_A\otimes\H_R \ .
\end{equation}
Note that the channel output $\M(\psi)$ is a state on $B$ and $R$ only.

Then it follows that 
\begin{equation}
    \max_{\psi\in\H_A\otimes\H_R}I(B:R)_{\M(\psi)}\le 2\pi L\langle H\rangle_{\M(\psi^*)}=:2\pi LE\ ,
\end{equation}
where $\psi^*$ denotes the maximizer of the LHS, and we denote the corresponding output state energy as $E:=\langle H\rangle_{\M(\psi^*)}$. The energy is finite as long as all the codewords used by Alice are regular states with finite energy. Now the information measure on the LHS is precisely the entanglement-assisted capacity $C_E(\M)$ as in~\eqref{eq:C_E}. The classical and quantum capacities are all dominated by $C_E$,
\begin{equation}\label{eq:bounded_capacities}
   C\le C_E\ , \  Q\le Q_E=C_E/2\ ,\  Q_0\le C_E\le 2\pi LE\ .
\end{equation}

Clearly, the classical and quantum capacities do not surpass the counterparts with entanglement assistance. Surprisingly, the zero-bits are also constrained by this Bekenstein bound: This follows from the fact the $C_E$ is equal to the amortized zero-bit capacity, which is the zero-bit capacity with catalytic use of a side noiseless quantum channel~\cite{hayden2012weak}.  The Bekenstein bound $2\pi LE$ thus holds for any quantum channels that satisfy the encoding and decoding constraints described above! 

Importantly, this bound pertains to both the encoder Alice and the decoder Bob: $L$ is the width of the overlapping region between them; and $\langle H\rangle_\rho$ measures the total energy Alice injects into the system with her encoding operations. Comparing it with the Bekenstein bound $\langle K_B\rangle=\beta E$ for the Unruh channel, now the zero-bit capacity is bounded because the encoder is restrained and she cannot prepare the Unruh modes as codewords. We stress that the bounded capacities~\eqref{eq:bounded_capacities} are not due to using a different bound, but rather that the channels under consideration have restricted encoding whereas the Unruh channel does not.  



Studying the Bekenstein bound for communication forces us to distinguish the restrictions on the encoder and the decoder. We have shown that this is actually a delicate issue that could determine if the Bekenstein bound indeed constrains the zero-bit capacity. 

Since we have considered the case of restraining only the decoder and the case of restraining both the encoder and the decoder, one could ask what if only the encoder is restrained. It turns out that it is still crucial to restrain the decoder for the Bekenstein bound to hold. We revisit an argument due to Page in Appendix~\ref{sec:counterexample}, which essentially shows that no Bekenstein bound could work otherwise. 

Therefore, we can conclude that it is \emph{necessary} for both the encoding and decoding to be constrained in order for the channel capacity to obey the Bekenstein bound that pertains to these restrictions. A precise characterisation of this class of 
restricted channels would be an interesting problem to explore.

\section{Concluding remarks}\label{sec:conclusion}

We have shown that in a quantum communication scenario, the Bekenstein bound can be evaded by zero-bits if only the receiver is spatially restrained. However, the bound can be restored when both the sender and the receiver are restrained. We close by making a few remarks on future directions.
\begin{itemize}
    \item \emph{Bekenstein bound for restricted channels.} It is unclear how to 
    formulate a general Bekenstein bound that applies to channels with restrictions on encoding and decoding. In particular, a prescription is needed to combine the spatial and energy constraints of the encoder and decoder in the bound. On the information-theoretic aspect, a major difficulty is that channel capacities are generally superadditive and we do not have a good understanding of this phenomenon nor effective tools to estimate the regularized quantities~\cite{wilde2013quantum}. Therefore, one feasible approach is to put a bound on the entanglement-assisted capacity $C_E$, which upper bounds almost all the other capacities of interest and admits a single letter formula. Hence, it is very useful to bound the mutual information between the decoder (decoding region) and the reference. Therefore, any generalisation of the Blanco-Casini bound on mutual information could be very useful.
    \item \emph{Bekenstein bound on the classical and quantum capacities.} It is natural to ask if the Bekenstein bound limits the classical capacity (and thus also the quantum capacity) for all channels that only constrain the decoder. This would be a direct generalisation for the Bekenstein bound on entropy. Consider an arbitrary quantum channel (in the Heisenberg picture) that embeds the operator subalgebra of some local region into the algebra of bounded operators on the QFT Hilbert space.  We would like to know if the regularized Holevo information is bounded by a Bekenstein bound that pertains to the decoding. In this case, a bound on $C_E$ is no longer available as we know $C_E$ of the Unruh channel is unbounded. Nonetheless, for any channel that outputs to a Rindler wedge, one can try to generalize the Unruh channel by lifting the restrictions we have made on the mode and one-particle codewords. This appears feasible because the Unruh channel can be treated as a Gaussian channel, so the formidable apparatus of Gaussian quantum information can be brought to bear to analyze the channel capacities~\cite{holevo1999capacity,holevo2001evaluating,giovannetti2004classical,caruso2006one}. 
    \item \emph{Operationality in QFT.} There is another fundamental issue with handling quantum information in QFT. Results in quantum information are usually proven in finite dimensions and it remains a nontrivial task to rigorously show that these results generalize to QFT.  The easy option is to set the technicalities aside and assume that the essential operational meaning of the entropic quantities calculated do carry over to QFT. This is plausible if the operations (preparations, channels, measurements) can be drawn freely from the local algebra.  Unfortunately, many operations that belong to the local algebra aren’t physical because they tend to violate causality~\cite{sorkin1993impossible}. There is only a limited subset of causal and local operations that are physically implementable~\cite{beckman2002measurability,borsten2021impossible,bostelmann2021impossible}. A full-fledged measurement theory is needed to ensure that the information-theoretic statements correspond to realizable operations. There has been an ongoing effort in this direction that builds the theory using detector models~\cite{polo2022detector} or within the framework of algebraic QFT~\cite{fewster2020quantum}. It would be useful to properly analyze quantum information protocols in QFT with the help of these new tools. 
    \item \emph{Zero-bit code.} The Unruh channel at high temperatures is particularly interesting because it has an extensive zero-bit capacity in contrast to the vanishing quantum capacity and negligible classical capacity.\footnote{In fact, the Unruh channel can be decomposed as a direct sum of optimal cloning channels~\cite{bradler2011infinite}, so the $1\to N$ cloning channels at large $N$ also share the same feature.  Another channel that shares this feature is the ``rocket channel''~\cite{smith2009extensive}, which is devised using random unitaries to showcase the non-additivity phenomenon. We thank Debbie Leung for pointing it out to us.} On the other hand, we know that the geometry preservation is poor at high temperatures so the Unruh channel itself is not a good zero-bit code. Therefore, one would need a good encoding to achieve the extensive zero-bit capacity for many parallel uses of this noisy Unruh channel. So far the noiseless classical channel remains the only channel for which we know an explicit capacity-achieving encoding~\cite{fawzi2013low}. It is of independent interest to find an explicit capacity-achieving code for the Unruh channel, which could also be physically illuminating.
    \item \emph{Zero-bits in holography.} Historically, the Bekenstein bound initiated the fruitful development of holographic entropy bounds that also universally appeal to gravity. These entropy bounds eventually led to the development of holographic dualities that deeply connect information and geometry. In this vein, we believe that our new twist to the old story could have further implications in quantum information and quantum gravity. For example, the fact that zero-bits are unconstrained by the Bekenstein bound should also extend to holographic bounds. After all, the Bekenstein bound was figured out by demanding any box of matter to be less entropic than the black hole area increase $\Delta A$ after it absorbs the matter. In fact, the same claim was made in holography~\cite{hayden2019learning}, where HP considered a scenario of two competing quantum Ryu-Takayanagi surfaces enveloping a mixed-state black hole. Then the zero-bit capacity of bulk-to-boundary holographic map is not bounded by the area difference between the two quantum Ryu-Takayanagi surfaces, whereas the quantum capacity is constrained. We believe it should be universally true in gravity that the zero-bits are not constrained by any holographic bound that conventionally applies to cbits/qubits.
\end{itemize}

\acknowledgments
We are grateful to Ahmed Almheiri, Raphael Bousso, Horacio Casini, Kfir Dolev, Henry Lin, Eduardo Martin-Martinez, Arvin Shahbazi-Moghaddam, Zhenbin Yang, and Shunyu Yao for their discussions and valuable comments. This work was supported by ARO (award W911NF2120214), CIFAR and the Simons Foundation.

\appendix

\section{No Bekenstein bound for $\alpha$-bits}\label{sec:alpha}

Hayden-Penington proposed a generalization of a zero-bit, called an $\alpha$-bit ($0\le\alpha\le 1$)~\cite{hayden2020approximate}. An $\alpha$-bit interpolates between a zero-bit ($\alpha=0$) and a qubit ($\alpha=1$). The intuition is that while a noisy channel might only be able to transmit a decent amount of noisy qubits, we can still make good use of them by distilling an $\alpha$-fraction of each qubit. Just like the zero-bits, the so-obtained $\alpha$-bits can be used as tighter communications resources in various tasks.\footnote{For instance, the $\alpha$-bit dense-coding resource inequality reads, $1\; \alpha$-bit $+ 1\; \mathrm{ebit}\ge (1+\alpha)\; \mathrm{cbits}$.}

Let us define $\alpha$-bits in the communication scenario. (cf. Section 3 of~\cite{hayden2020approximate} for the precise definition.) Recall that the zero-bit capacity $Q_0(\M)$ characterizes how much information the channel can transmit if we only require Bob to decode a two-dimensional (or any constant) subspace. The \emph{$\alpha$-bit capacity} $Q_\alpha(\M)$ is defined similarly by requiring Bob to decode, with sufficiently small error at large enough $n$, a $k^{n\alpha}$-dimensional subspace of the $k^n$-dimensional input message transmitted by $\M^{\otimes n}$.  We then say that a channel $\M$ transmits a $Q_\alpha(\M)$ amount of $\alpha$-bits (asymptotically).

Like the zero-bit capacity, the $\alpha$-bit capacity of a (conjugate) degradable channel $\M:\mathcal{P}(\H_A)\to\mathcal{P}(\H_B)$ admits a single-letter formula,
\begin{equation}\label{eq:abit_capacity}
    Q_\alpha(\M) = \sup_{\psi\in\H_A\otimes\H_{A'}}\left[ \min\left(\frac{1}{1+\alpha}I(A':B)_{\M(\psi)},\frac{1}{\alpha}I(A\rangle B)_{\M(\psi)}\right)\right]\ .
\end{equation}
From the continuity of the $\alpha$-bit capacity in $\alpha$, we expect a violation of the Bekenstein bound for some, if not all, values of $\alpha$'s.  

For our Unruh channel, the two information measures involved in~\eqref{eq:abit_capacity} are evaluated for $C_E(\N)$ and $Q(\N)$ respectively. They are both maximized at the maximally entangled state, so it is straightforward to evaluate $Q_\alpha(\N)$. We obtain
\begin{equation}\label{eq:abit_capacity_N}
   Q_\alpha(\N)= \begin{cases}
        \frac{1}{1+\alpha}(Q(\N)+\log_2 d),\quad& \text{when}\quad \alpha\le Q(N)/\log_2 d\\
        Q(\N)/\alpha,& \text{when}\quad \alpha > Q(N)/\log_2 d
    \end{cases}
\end{equation}
where $Q(\N)$ is given in \eqref{eq:qc}. 

\begin{figure}
    \centering
    \includegraphics[width=.65\linewidth]{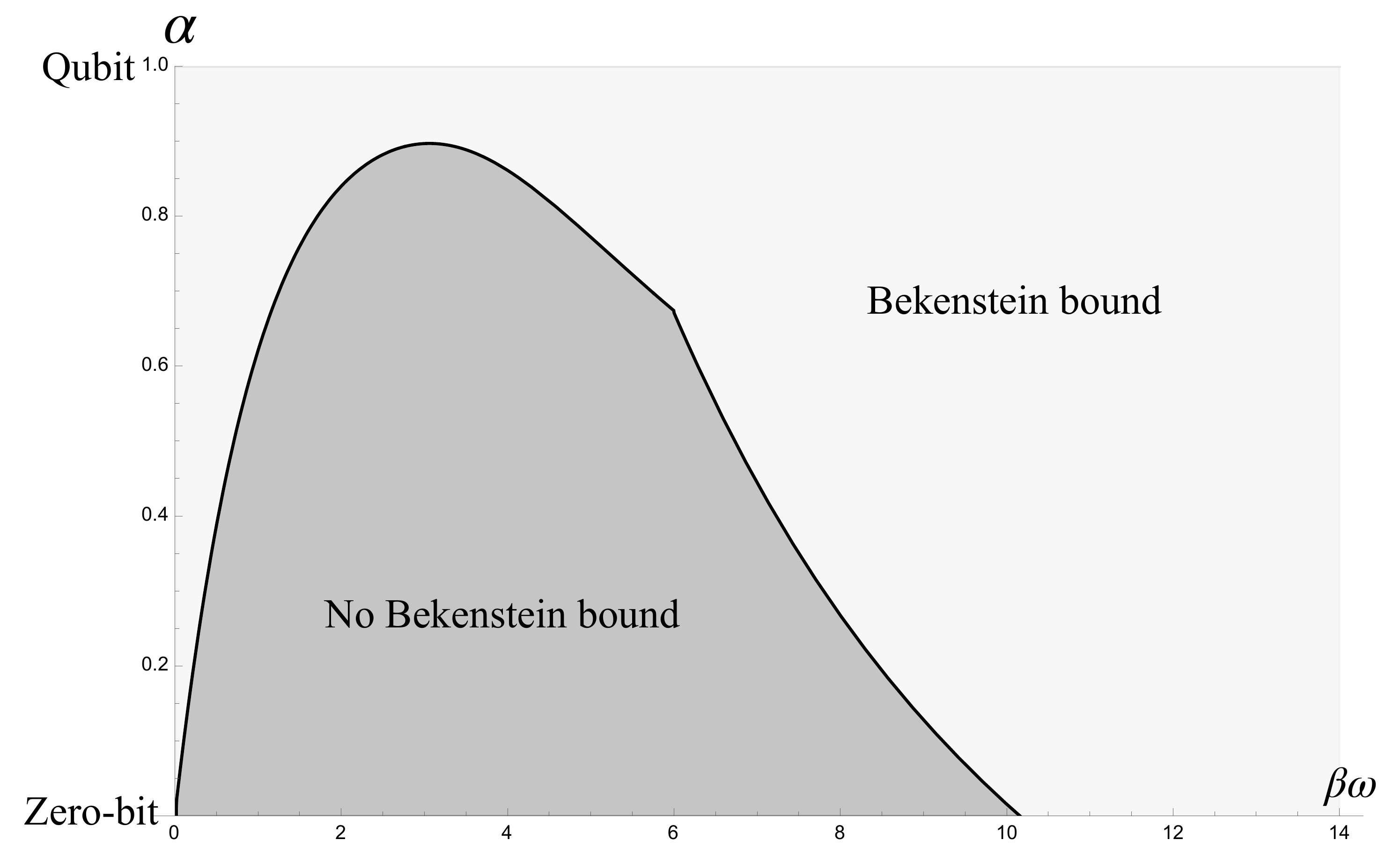}
    \caption{\textbf{No Bekenstein bound for $\alpha$-bits.} This phase diagram shows which $\alpha$-bits transmitted by the Unruh channel obey the Bekenstein bound as one varies the temperature (or frequency). The shaded region indicates transcending the bound. We see that for a large enough $\alpha$, the Bekenstein bound always holds at any temperature. }
    \label{fig:abits}
\end{figure}

It is natural to ask for which $\alpha$ values there is a violation of the Bekenstein bound $\langle K_B\rangle=\beta E$. Given a fixed $\beta\omega$, we find that there is a violation for
\begin{equation}\label{eq:phase}
    \alpha\le \begin{cases}
      Q(\N)/\beta E \quad& \text{when}\quad  \log_2 d\ge\beta E\\
      (Q(\N)+\log_2 d)/\beta E -1\quad& \text{when}\quad  \log_2 d <\beta E
    \end{cases}\ .
\end{equation}
As a sanity check, we note that there is no lower bound for the range of Bekenstein-violating $\alpha$ values, so the zero-bits always violate the bound at some temperature. On the other hand, the upper bounds are strictly less than $1$ because $Q(\N)<\beta E$, meaning that qubits obey the Bekenstein bound. Also, the maximal $\alpha$ that satisfies \eqref{eq:phase} can be arbitrarily close to $1$ at a sufficiently large $d$, because in this limit $Q(\N)$ tends to $\beta E$ from below. In Fig.~\ref{fig:abits}, we plotted a ``phase diagram'' that illustrates \eqref{eq:phase} for $d=2^6$.

\section{Page's proposal revisited}\label{sec:counterexample}

Another situation where Casini's entropy bound $\langle K_B\rangle$ applies to the accessible information is when the information is encoded by local unitaries. This is the scenario that Page considered~\cite{page2008defining}. Page let Alice encode her message only via local unitaries supported over the region $B$. Then Page suggested considering whether the entropy of $S(\bar\rho)$, where $\bar\rho$ is the global mixture of the codewords, is Bekenstein-bounded by the width $R$ of the region, times the energy $E:=\langle H\rangle_{\bar\rho}$. 

Unfortunately, Page showed that any such Bekenstein bound can be violated by engineering a mixture between the vacuum state and any excited state $\ket{\psi}=U\ket{\Omega}$, even if $U$ is locally supported in region $B$. The mixed state reads $\bar\rho=(1-p)\Omega+p \psi$. We do not assume that $\psi$ is orthogonal to $\Omega$ so we let the state overlap be $r=\tr\ \Omega\psi$. Suppose that the Hamiltonian of the theory $H$ is renormalized such that $H\ket{\Omega}=0$. Then the energy of $\rho$ reads $E=\langle H\rangle_{\bar\rho} = p\bra{\psi}H\ket{\psi}$. Hence, the Bekenstein bound can be arbitrarily small and scales \emph{linearly} with $p$.

Consider now the entropy of $\bar\rho$, which has eigenvalues $\{q, 1-q\}$ and $q=p(1-p)(1-r)$. The entropy of $\bar\rho$ is given by the binary entropy function $h(q):=-q\log q-(1-q)\log(1-q)$. Note that at small $p$, $q$ scales linearly with $p$ and $h(q)$ scales like $-q\log(q)$. Therefore, the gradient of the entropy at small $p$ goes like $\O(-\log p)$, and for sufficiently small $p$ it overwhelms the Bekenstein bound which has a finite gradient $\bra{\psi}H\ket{\psi}R$. 

In the operational context of distinguishing an ensemble of states created by local unitaries, entropy is an upper bound on the accessible information. The bound is not tight when the states are nonorthogonal. This gap leaves the possibility that the Bekenstein bound could still apply to the accessible information instead. We now show that the same counterexample also forbids this possibility. Namely, when $p$ is small enough, the accessible information from the ensemble $\{(1-p,\Omega), (p,\psi)\}$ surpasses the Bekenstein bound $RE$. 

The violation essentially boils down to the same divergent gradient at $p=0$ of the accessible information. For an ensemble of two pure states parameterized by the mixing probability $p$ and the state overlap $r$, the optimal measurement is the Holevo-Helstrom measurement, and the accessible information $I_\mathrm{acc}(p,r)$ has an explicit expression (see equation (23) in~\cite{levitin1995optimal}). It turns out that the gradient $\partial_p I_\mathrm{acc}(p,r)$ has the same logarithmic divergence near $p=0$, so the violation follows.

What kind of information is captured by this (global) entropy that fails to admit a Bekenstein bound? Note that this entropy is the (global) Holevo information of the ensemble because all the states are pure. Therefore, $S(\bar\rho)$ upper bounds the globally accessible information $I_\mathrm{acc}(BB^c,\rho)$, and Page's proposal amounts to finding a Bekenstein bound for the globally accessible information. For example, suppose that Alice can encode her message optimally so that the ensemble $\{\rho^x\}$ is an orthonormal set. Then the globally accessible information is exactly given by $S(\bar\rho)$. 

Let us understand why this proposal fails. We claim the reason is that the decoding is not spatially constrained. The globally accessible information presumes that Bob's measurements are not restricted to acting on $B$ alone. While it is true that Alice's local unitary encoding prevents any leakage of information to $B^c$, meaning Bob can learn nothing from $B^c$ on its own, Bob can nonetheless learn more by accessing $B^c$ \emph{in addition} to $B$. That is because he can perform nonlocal measurements jointly on $BB^c$, and the entanglement in the vacuum allows Bob to probe the different correlations created by Alice's local encoding. In contrast, Casini's proposal~\eqref{eq:casini} works because the state distinguishing task is operationally confined to the region $B$, and the relative entropy depends only on the reduced states on $B$. 

We can ``fix'' Page's proposal by restricting Bob's measurement to the region $B$ only. Then the bound~\eqref{eq:bound_on_chi} can be applied to the information locally accessible from $B$. Moreover, since the encodings are unitary, $-\sum_x p_x S(\rho^x_B)=-\sum_x p_x S(\Omega_B)=-S(\Omega_B)$. Therefore, \eqref{eq:bound_on_chi} is equivalent to the vacuum-subtracted bound of Casini~\eqref{eq:casini}. We thus have
\begin{equation}
I_\mathrm{acc}(B,\rho)\le \sum_x p_x S(\rho^x_B||\Omega_B)= \delta \langle K_B\rangle_{\bar\rho_B}.
\end{equation}
Our result in Section~\ref{sec:encoder} shows that one can also restrain Bob in such a way that his decoding region overlaps with Alice's encoding region in order for the Blanco-Casini bound to apply.

\section{Some estimates}\label{sec:estimates}
\begin{figure}
     \centering
     \begin{subfigure}[b]{0.51\textwidth}
         \centering
         \includegraphics[width=\textwidth]{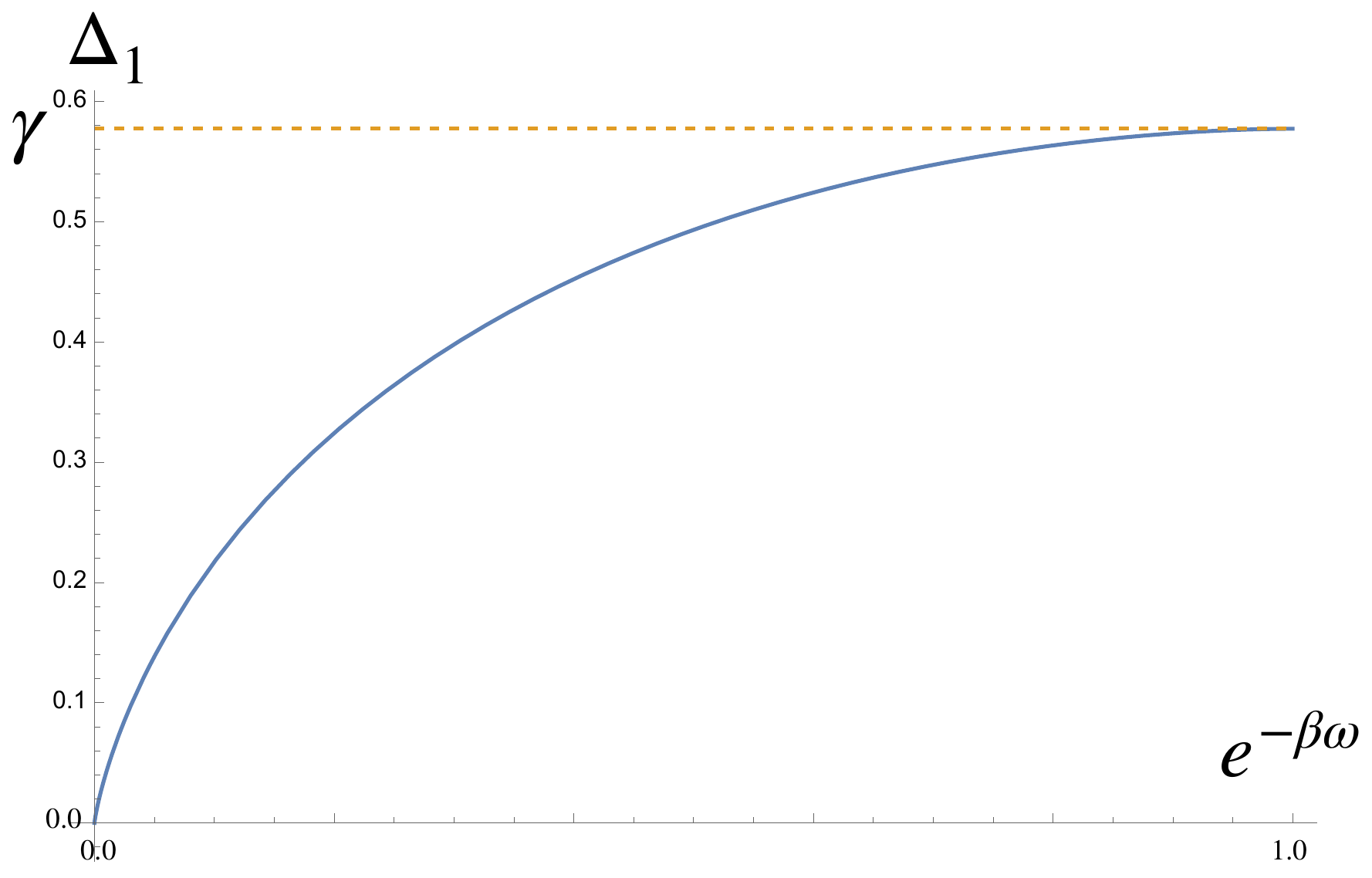}
         \caption{\centering $\Delta_1:=H(1)-H_0(1)$}
         \label{fig:estimate1}
     \end{subfigure}
     \hfill
     \begin{subfigure}[b]{0.48\textwidth}
         \centering
         \includegraphics[width=\textwidth]{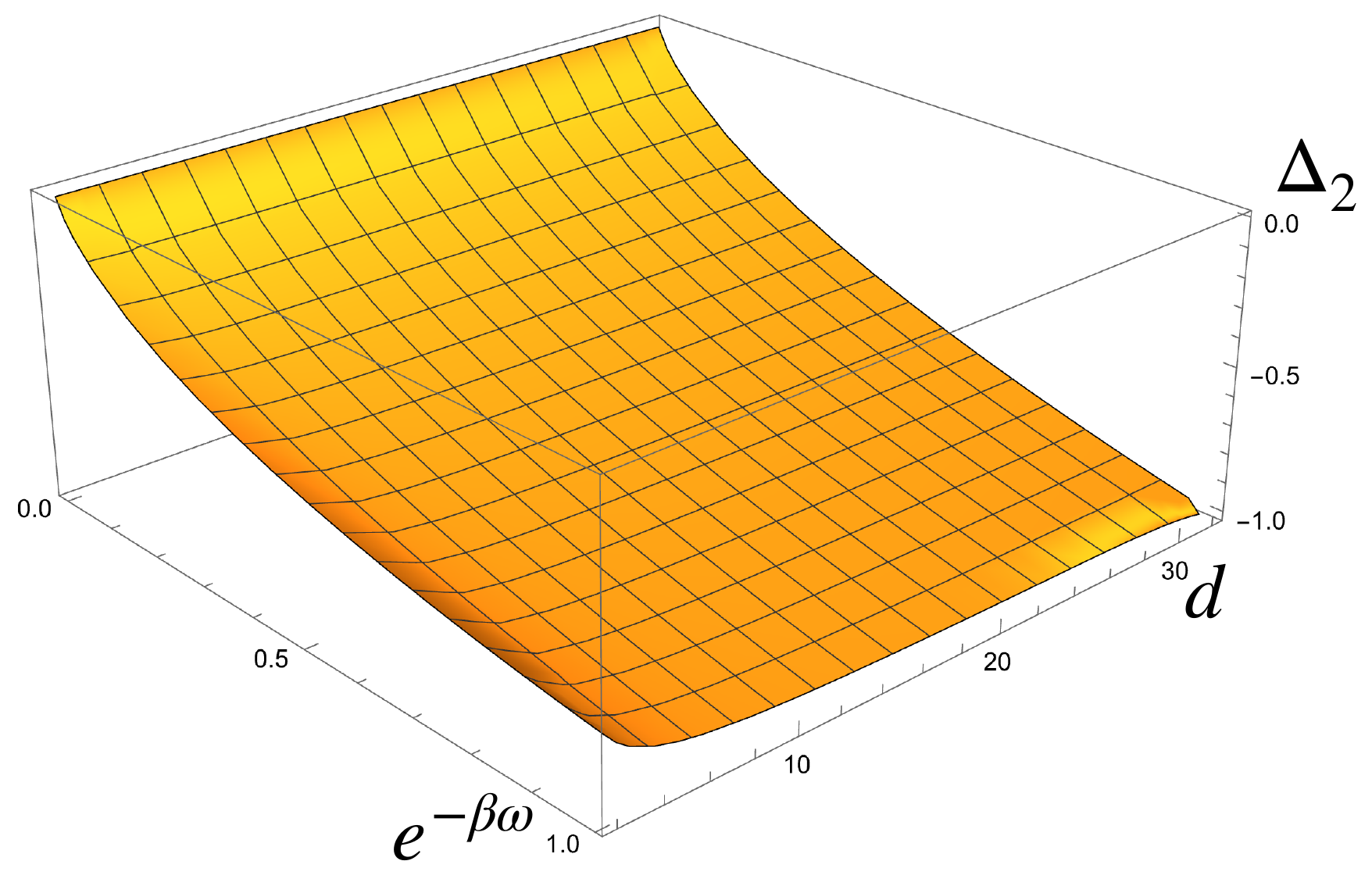}
         \caption{\centering $\Delta_2:=H_0(d) - H^c(d)$}
         \label{fig:estimate2}
     \end{subfigure}
        \caption{Plots in support of \eqref{eq:estimate1} and \eqref{eq:estimate2}.}
        \label{fig:estimates}
\end{figure}

We give some details of some estimates that are used in the main text.  Let us collect some entropy formulas which can be found in BHP~\cite{bradler2012quantum}.
\begin{equation}
    H_0(d):=S(\Omega) = d\left(\frac{ \beta \omega}{e^{\beta \omega}-1}-\log(1-e^{-\beta \omega})\right)\ ,
\end{equation}
\begin{multline}
    H(d):=S(\N(\pi))=\log d-(d+1)\log(1-e^{-\beta\omega})+(1+d)\frac{ \beta\omega e^{-\beta\omega}}{1-e^{-\beta\omega}}\\-\frac{(1-e^{-\beta\omega})^{d+1}}{d}\sum_{k=1}^\infty\binom{d+k-1}{k}k\log k e^{-\beta\omega(k-1)}\ ,
\end{multline}
 
\begin{multline}
    H^c(d):=S(\N^c(\pi))=\log d-(d+1)\log(1-e^{-\beta\omega})+(1+d)\frac{ \beta\omega e^{-\beta\omega}}{1-e^{-\beta\omega}}\\-\frac{(1-e^{-\beta\omega})^{d+1}}{d}\sum_{k=1}^\infty\binom{d+k-1}{k}k\log(d+k-1) e^{-\beta\omega(k-1)}\ .
\end{multline}
Let us first justify the claim used to estimate the value of classical capacity
\begin{equation}\label{eq:estimate1}
    \gamma  \ge H(1)-H_0(1) \ge 0\ .
\end{equation}
We have 
\begin{multline}
    H(1)-H_0(1) = - \log(1-e^{-\beta\omega})+\frac{ \beta\omega e^{-\beta\omega}}{1-e^{-\beta\omega}}-(1-e^{-\beta\omega})^2\sum_{k=1}^\infty k\log k e^{-\beta\omega(k-1)}\\
    =h(e^{-\beta\omega})/(1-e^{-\beta\omega}) + e^{\beta\omega}(1-e^{-\beta\omega})^2 \partial_s\mathrm{Li_s(e^{-\beta\omega})}|_{s=-1}\ ,
\end{multline}
where $h(\ \cdot \ )$ is the binary entropy function and $\mathrm{Li_s}$ is the poly-logarithmic function of order~$s$. We can use \textsc{Mathematica} to verify that this function of $\beta\omega$ satisfies~\eqref{eq:estimate1}. We give a plot of it in Fig.~\ref{fig:estimate1}. 

We also used the following to estimate the value of quantum capacity,
\begin{equation}\label{eq:estimate2}
    -1\le H_0(d) - H^c(d)\le 0\ .
\end{equation} 
Note that the $\le 0$ part essentially follows the fact that the quantum capacity must be smaller than the quantum capacity. Unlike~\eqref{eq:estimate1}, $H_0(d) - H^c(d)$ depends on the dimension. Since we cannot analytically evaluate the sum in $H^c(d)$ for all $d$, we provide some numerical evidences in Fig.~\ref{fig:estimate2} that~\eqref{eq:estimate2} holds. The exact lower bound $-1$ is not important to establish our main results.

\section{Zero-bit dense coding}\label{sec:densecoding}
We first remind the readers of the standard dense coding protocol which helps send $2\log d$ cbits with a qudit or a channel that has $\log d$ quantum capacity. Alice and Bob share a $d^2$-dimensional Bell state $\Psi$. Alice acts following local unitaries 
\begin{equation}
    X :=\sum_{0\le j <d}\ketbra{j+1 \ \text{mod}\ d}{j},\quad Z:= \sum_{0\le j <d}e^{i 2\pi j/d}\ketbra{j}{j}\ 
\end{equation}
on her share of the Bell state depending on her $2\log d$-bits message. She divides her message string into two equal parts, and we can label each message with $xy$, where $0\le x,y <d$ labels the message string. Given a message $xy$, Alice prepares the orthonormal codewords as follows
\begin{equation}
   \ket{\Psi_{xy}} :=Z^x X^y \ket{\Psi}\ .
\end{equation}
She then sends her share (a qudit) to Bob, and Bob can decode the cbits from a $d^2$ orthonormal set of Bell states.\\

The zero-bit dense coding protocol of Hayden-Penington~\cite{hayden2020approximate} can send cbits at a rate of $\log_2 d$ with $\log_2 d$ zero-bits. In the one-shot setting, the protocol can help send $\log d + 1 $ cbits with some error $\delta(\eps)$ using a zero-dit with some error $\eps$. 

Alice now has a zero-dit instead of a qudit at her disposal. She encodes her message of $\log_2 d$ bits entirely in $x$, plus one more bit $y\in \{0,1\}$.
\begin{equation}
   \ket{\Psi_x} = Z^x X^y\ket{\Psi} = \frac{1}{\sqrt{d}}\sum_{0\le j <d} e^{i 2\pi x j/d}\ket{j}_{B'}\ket{j+y}_A. 
\end{equation}
Alice then sends her share to Bob as a zero-dit, which only allows Bob to decode any two dimensional subspace. Had Bob chosen to measure his own share of the Bell pair in the computational basis, he can then figure out the $y$ bit. More explicitly, let the Stinespring dilation of Alice's encoding and channel be $U^{A\to BE}$. For any state in $\ket{\chi_j}_A\in\mathrm{span}\{\ket{j}_A,\ket{j+1}_A\}$, Bob applies a corresponding decoding isometry $V_j^{B\to AE'}$
\begin{equation}
    V_j^{B\to AE'}U^{A\to BE}\ket{\chi_j}_A\approx \ket{\chi_j}_A\otimes \ket{\eta}_{EE'}
\end{equation}
where the approximation error depends on the $\eps$ of the zero-dit; and the environment state $\ket{\eta}_{EE'}$ can be set independent of $j$. 
The purification of the channel output is 
\begin{equation}
    \frac{1}{\sqrt{d}}\sum_j e^{i 2\pi x j/d}\ket{j}_{B'}U^{A\to BE}\ket{\chi_j}_A\ .
\end{equation}
Bob applies the following isometry on the above state
\begin{equation}
    V=\sum_j\ketbra{j}{j}\otimes V_j^{B\to AE'}
\end{equation}
and we obtain
\begin{equation}
    \frac{1}{\sqrt{d}}\sum_j e^{i 2\pi x j/d}\ket{j}_{B'}V_j^{B\to AE'}U^{A\to BE}\ket{\chi_j}_A \approx \left(\frac{1}{\sqrt{d}}\sum_j e^{i 2\pi x j/d}\ket{j}_{B'}\ket{\chi_j}_A\right)\otimes\ket{\eta}_{EE'}
\end{equation}
For $\ket{\chi_j}_A\in\{\ket{j}_A,\ket{j+1}_A\}$, we obtain Alice's codewords and Bob can measure and decode the $(\log d +1)$-bits classical message.

See Theorem 5 in~\cite{hayden2020approximate} for the proof that this protocol achieves the entanglement-assisted capacity of $\log_2 d$ with vanishing error.

\bibliographystyle{JHEP}
\bibliography{zero}

\end{document}